\newcommand{\subtil}[1]{\makebox[0mm][l]{\raisebox{-3mm}[3mm][-2mm]{$\
\tilde{}$}}#1}
\newcommand{\One}{{\bf 1}}
\newcommand{\ag}{\alpha}
\newcommand{\bg}{\beta}
\newcommand{\cg}{\gamma}
\newcommand{\dg}{\delta}
\newcommand{\sg}{\sigma}
\newcommand{\lam}{\lambda}
\newcommand{\Lam}{\Lambda}
\newcommand{\tilep}{\tilde{\varepsilon}}
\newcommand{\suchthat}{\ni}
\newcommand{\di}{\partial}
\newcommand{\be}{\begin{equation}}
\newcommand{\ee}{\end{equation}}
\newcommand{\bearr}{\begin{eqnarray}}
\newcommand{\eearr}{\end{eqnarray}}
\newcommand{\Sg}{\Sigma}
\newcommand{\Sgauge}{\Sigma_{i\,0a}}
\newcommand{\m}{\mu}
\newcommand{\n}{\nu}
\newcommand{\eg}{\epsilon}
\newcommand{\tilE}{\tilde{E}}
\newcommand{\tilB}{\tilde{B}}
\newcommand{\tilw}{\tilde{w}}
\newcommand{\tilDg}{\tilde{\Delta}}

\newcommand{\implies}{\Rightarrow}
\newcommand{\genSD}{\tilB^{i\,a} - \phi^{ij}\tilE^a_j}

\newcommand{\exchange}{\leftrightarrow}
\newcounter{eqnum_mem}
\newcounter{gSDnum}
\newcounter{Gaussnum}
\newcommand{\Dg}{\Delta}
\newcommand{\Dgthree}{{}_{\scriptscriptstyle 3}\Delta}
\newcommand{\Real}{{\bf R}}
\newcommand{\Complex}{{\bf C}}
\newcommand{\QED}{\rule{1mm}{3mm}}
\newcommand{\calB}{{\cal B}}
\newcommand{\calS}{{\cal S}}
\newcommand{\calP}{{\cal P}}
\newcommand{\caltilG}{\tilde{\cal G}}
\newcommand{\caltilC}{\tilde{\cal C}}
\newcommand{\caltilV}{\tilde{\cal V}}
\newcommand{\caltilS}{\:\tilde{\!\!\tilde{\cal S}}}
\newcommand{\tilpi}{\tilde{\pi}}
\newcommand{\Egj}{\tilE_{(\cg,j)}{}}
\newcommand{\loopstate}{\langle A|\cg,j\rangle}
\newcommand{\DiffM}{\mbox{\em Diff}_0(M)}
\newcommand{\fun}{\mbox{\tiny fun}}
\newcommand{\adj}{\mbox{\tiny adj}}
\documentstyle[12pt,psfig]{article}
\begin{document}

\title{New constraints for canonical general relativity}
\author{Michael P. Reisenberger\\
	Institute for Theoretical Physics\\
	University of Utrecht, P. O. Box 80 006\\
	3508 TA Utrecht, The Netherlands}
\maketitle

\begin{abstract}
Ashtekar's canonical theory of classical complex Euclidean GR (no Lorentzian
reality conditions) is found to be invariant under the full algebra of
infinitesimal 4-diffeomorphisms, but non-invariant under some finite proper
4-diffeos when the densitized dreibein, $\tilE^a_i$, is degenerate. The
breakdown of 4-diffeo invariance appears to be due to the inability of the
Ashtekar Hamiltonian to generate births and deaths of $\tilE$ flux loops
(leaving open the possibility that a new `causality condition' forbidding
the birth of flux loops might justify the non-invariance of the theory).

A fully 4-diffeo invariant canonical theory in Ashtekar's variables, derived
from Plebanski's action, is found to have constraints that are stronger than
Ashtekar's for $rank\tilE < 2$. The corresponding Hamiltonian generates
births and deaths of $\tilE$ flux loops.

It is argued that this implies a finite amplitude for births and deaths of
loops in the physical states of quantum GR in the loop representation,
thus modifying this (partly defined) theory substantially.

Some of the new constraints are second class, leading to difficulties in
quantization in the connection representation. This problem might be
overcome in a very nice way by transforming to
the classical loop variables, or the `Faraday line' variables of Newman and
Rovelli, and then solving the offending constraints.

Note that, though motivated by quantum considerations, the present paper is
classical in substance.
\end{abstract}

\section{Introduction}

In 1986 Ashtekar presented a new set of canonical variables for general
relativity \cite{Ashtekar86},
\cite{Ashtekar87}, namely the spatial self-dual spin connection $A^i_a$ and,
canonically conjugate
to it, the densitized triad $\tilde{E}^a_i$. The constraints on the physical
phase space of the
ADM formulation \cite{ADM62} were translated, by canonical transformation, into
the new variables,
and were found to be simple polynomials.

On the ADM phase space the $3\times 3$ matrix $\tilde{E}^a_i$ is invertible.
Ashtekar dropped this requirement and defined his canonical theory on the
larger phase space consisting of all pairs of
fields $(A^i_a, \tilE^a_i)$,\footnote{
Of course there are differentiability conditions, and, in the asymptotically
flat case, fall off conditions.}
including ones in which $\tilE^a_i$ is degenerate. As the constraints on the
degenerate part of the phase space he simply used the same polynomial
expressions that he had found in the non-degenerate, ADM, sector. Certain
degenerate field configurations (which are not gauge equivalent to
non-degenerate ones) solve Ashtekar's constraints, so the physical phase space
of his theory is also somewhat bigger than that of the ADM theory.

The question now arises: is this extension of the ADM theory still
4-diffeomorphism invariant? The answer turns out to be that it is invariant
under {\em infinitesimal} 4-diffeos, but for some degenerate solutions there
are finite 4-diffeos (connected with the identity) that do not map them to
solutions.

A nice way to construct a 4-diffeo invariant canonical theory is to derive it
from a manifestly invariant action.
In the present paper we use the Plebanski action for GR
\cite{Plebanski77} as our starting point, and derive the full set of
corresponding constraints
on the Ashtekar variables from it, paying special attention to the case
of degenerate $\tilE$.\footnote{
The constraints have been derived from the Plebanski action before (see
\cite{CDJM}) but the case of degenerate $\tilE$ was not treated.}
%
The Plebanski action leads to classical field equations which are equivalent to
the Einstein equations when the latter are defined, but which are themselves
defined on a larger class of spacetime geometries. In particular, the action is
defined
on geometries the spatial cross sections of which have degenerate $\tilde{E}$.
Via a $3+1$ decomposition
of spacetime, and a Legendre transformation, the Plebanski action leads
naturally to a canonical theory in terms of Ashtekar's
variables, with no {\em a priori} restrictions on the rank of
$\tilde{E}$.\footnote{%
The Plebanski action does not provide the only 4-diffeo invariant extension
of GR to degenerate geometries. A distinct theory is defined by the
action
\be
I' = \int \hat{e}_I{}^\m \hat{e}_J{}^\n F^{+IJ}_{\m\n}\ d^4x
\ee
where $\hat{e}_I{}^\m = det[e^J{}_\n]^{\frac{1}{2}} e_I{}^\m$ is the weight
$\frac{1}{2}$ densitized vierbein, and $F^{+IJ}$ is the curvature of the
self-dual connection $A^{+IJ}$.
This action, which is a hybrid of the Samuel-Jacobson-Smolin action
\cite{Samuel87}\cite{Jacobson_Smolin87}\cite{Jacobson_Smolin88a} and
Deser's action \cite{Deser70}, was suggested
to the author by Ingemar Bengtsson and is implicit in \cite{Bengtsson89}. The
corresponding canonical theory
appears to share with that derived from the Plebanski action, the crucial
feature that the Hamiltonian generates births and deaths of $\tilE$
flux loops. However, the action $I'$ will not be discussed further in this
paper. }

The Lorentzian reality conditions are {\em not} imposed. The fields are defined
so that when they are all real the Euclidean theory is obtained. The paper
therefore treats complex GR, in a Euclidean notation.\footnote{
Real Lorentzian GR is a specialization of complex Euclidean GR obtained by
imposing the Lorentzian reality conditions. To recover real Euclidean GR one
simply requires that $A$ and $\tilE$ are both real. These algebraic conditions
are preserved by the evolution. To obtain real Lorentzian GR one still requires
that $\tilE^a_i$ is real, at least up to internal $SO(3)$ gauge transformations
so that the densitized metric, $\tilE^{i\,a}\tilE^b_i$, is real, but instead of
requiring $A$ to be real one must impose a differential constraint that
ensures that the reality of $\tilE$ (up to $SO(3)$ gauge) is preserved in time.
See \cite{Ashtekar_Tate}. This differential constraint, which is not dealt with
in the present paper, changes the theory profoundly. Thus results of this paper
can only be applied straightforwardly to Euclidean GR.}

The constraints are found to be
\setcounter{Gaussnum}{\value{equation}}
\begin{eqnarray}
D_a \tilde{E}^a_i & = & 0 \label{Gauss_law} \\
\setcounter{gSDnum}{\value{equation}}
\exists \phi^{ij} \mbox{trace free, symmetric} & \suchthat & \tilde{B}^{i\,a} -
\phi^{ij} \tilde{E}^a_j = 0.  \label{generalized_SD}
\end{eqnarray}
$D$ is the covariant derivative defined by the self-dual connection $A$, and
$\tilde{B}^{i\,a}$ is (twice) the magnetic field of that connection. Latin
indices from
the
beginning of the alphabet, $a, b, ...$ are external vector indices, while those
from the middle,
$i, j, ...$ are internal $SO(3)$ indices. $\tilde{\ }$ on top of a field
variable indicates that the field is a 3-space density of weight $1$ (like the
determinant of the co-triad $E^i_a$).

(\ref{generalized_SD}) implies Ashtekar's vector and scalar constraints
\begin{eqnarray}
\tilde{B}^{i\,a}\tilde{E}_i^b \epsilon_{abc} & = & 0 \label{vector} \\
\tilde{B}^{i\,a}\tilde{E}_j^b\tilde{E}_k^c \epsilon_{abc} \epsilon_i{}^{jk} & =
& 0, \label{scalar}
\end{eqnarray}
but the converse is not true when $rank\,\tilde{E}<2$. For this subset of the
degenerate field
configurations the constraints (\ref{Gauss_law}) and (\ref{generalized_SD}) are
stronger than Ashtekar's constraints (\ref{Gauss_law}), (\ref{vector}),
and (\ref{scalar}). Consequently,
the new Hamiltonian, which is a linear combination of (\ref{Gauss_law}) and
(\ref{generalized_SD}),
generates a larger class of possible (gauge) evolutions than does the Ashtekar
Hamiltonian.\footnote{%
In \cite{CDJ89} Cappovilla, Dell and Jacobson found something similar to
(\ref{generalized_SD}). They noted that $\tilE^a_i =
\phi^{-1}_{ij}\tilB^{j\,a}$,
with $\phi^{ij}$ an invertible, traceless, symmetric matrix,
is the general solution to Ashtekar's constraints (\ref{vector}) and
(\ref{scalar})
when $\tilE$ and $\tilB$ are non-degenerate. In \cite{CDJ} they present an
action
which makes this equation (and the Gauss law (\ref{Gauss_law})) the fundamental
canonical
constraints. Note, however, that this theory is not equivalent to Plebanski's
theory,
of which (\ref{generalized_SD}) and (\ref{Gauss_law}) are the constraints.
Their theory excludes solutions with $\tilE \neq 0$, $\tilB = 0$ (such as flat
space-time)
which Plebanski's theory does not.}

One might think that the difference between Ashtekar's constraints and the new
constraints
is not significant, since (\ref{vector}) and (\ref{scalar}) imply
(\ref{generalized_SD}) on
generic $(A,\tilde{E})$ field configurations. I will now argue, somewhat
heuristically, that (\ref{generalized_SD})
leads to a profoundly different quantum theory from that of (\ref{vector}) and
(\ref{scalar}). At least if that theory is formulated via `loop quantization'
\cite{Gambini86} \cite{RSloops90}. I should emphasize, however, that the
following
arguments, which touch on quantum theory, are strictly for motivation.
The results of the paper are entirely classical and do not depend on the
following arguments.

\begin{figure}
\centerline{\psfig{figure=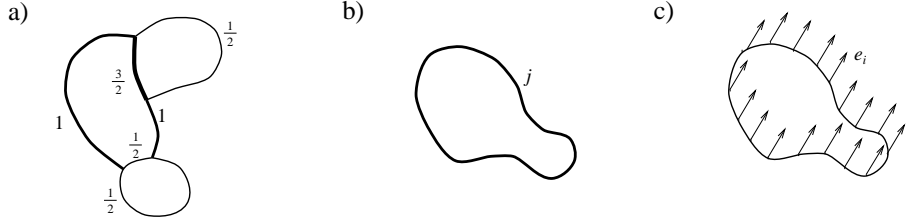,height=5cm}}
\caption[xxx]{Panel a) illustrates a graph basis state. Graph basis bras,
$\langle \Gamma|$, are linear combinations of loop basis bras that span
the space of solutions to the Mandelstam constraints. That is, the graph
amplitudes $\Psi[\Gamma]$ are independent and completely parametrize
$SU(2)$ gauge invariant states $|\Psi\rangle$.
\newline\newline
Panel b) shows a graph, $\cg_{(j)}$, consisting of a single loop carrying spin
$j$. When $j = \frac{1}{2}$ $\langle \cg_{(j)}|$ is just the loop basis bra
of the same loop. This loop basis bra can be represented by
$tr\,H^{(\frac{1}{2})}[A,\cg]$,
the trace of the spin $\frac{1}{2}$ holonomy around $\cg$, with the amplitude
for the loop
in a state $|\Psi\rangle$ given by the loop transform
\newline\newline
$\ \ \ \Psi[\cg] = \langle \cg|\Psi\rangle =
\int d\m[A]\:(tr\,H^{(\frac{1}{2})}[A,\cg])^*\: \Psi[A]$.
\newline\newline
($\int d\m[A]$ is an integral over $SU(2)$ connections defined on the relevant
class of
functionals of these connections). $\langle \cg_{(j)}|$ is then represented by
$tr\,H^{(j)}[A,\cg]$.
In other words, $\langle\cg_{(j)}|$ for $j>\frac{1}{2}$ is just like the loop
basis bra
$\langle\cg_{(\frac{1}{2})}|$ except the spin $j$ holonomy replaces the spin
$\frac{1}{2}$ holonomy in the loop transform.

When $d\m[A]$ is taken to be the ``induced Haar measure'' (see
\protect\cite{Ash_Lew})
graph basis states are orthonormal in the inner product $\langle
\theta|\varphi\rangle = \int d\m[A]\: \theta^*\varphi$, so we can think of
$\Psi[\Gamma]$ as the coefficients in an expansion of the state $|\Psi\rangle$
on graph {\em ket} states $|\Gamma\rangle$, with $\langle A |\cg_{(j)}\rangle
= tr\,H^{(j)}[A,\cg]$.
\newline\newline
Panel c) shows an $\tilE$ flux loop. In the classical limit $\hbar\rightarrow
0$, $j \sim O(1/\hbar)$ $\ |\cg_{(j)}\rangle$ represents such a flux loop.}
\label{graph_basis}
\end{figure}

 The argument is most easily described in
the context of a variant of loop quantization, which might be called
`graph quantization'
\cite{Baez94}, \cite{discarea2}, \cite{Reisenberger94}). In graph quantization
the states are
superpositions of `graph basis states', $|\Gamma\rangle$, associated with
graphs, $\Gamma$, in 3-space whose edges and vertices are colored by non-unit
irreducible representations ($j\in \{\frac{1}{2}, 1,\frac{3}{2}, 2, ...\}$) of
$SU(2)$. (See Fig. \ref{graph_basis}).

A particularly simple class of graphs are loops $\cg$, without intersections,
carrying spin $j$. In the classical limit the corresponding basis
states $|\cg,j\rangle$ represent isolated Faraday lines, or `flux loops':
\be	\label{fluxline1}
\tilE^a_i = \pm e_i \int_\cg \dg^3 (x-z) dz^a,
\ee
where $e_i$ has magnitude $j\hbar$ and is covariantly constant along $\cg$.
(See Appendix \ref{classical_loop} for a proof and caveats).

Notice that on $\gamma$ $\ rank\tilE = 1$, and off $\cg$ $\ \tilE = 0$, so we
expect the evolution of these field configurations in the new canonical theory
to differ from that in Ashtekar's theory. Indeed in Ashtekar's theory
(\ref{fluxline1}) can only evolve by 3-diffeomorphisms, i.e. they can move
around in space, while in the new theory loops of $\tilE$ flux can appear from,
and disappear into `vacuum', $\tilE = 0$. Appearances and disappearances of
$\tilE$ flux loops will be referred to, respectively, as `births' and `deaths'.

The lack of births and deaths in the classical Ashtekar theory seems to be
mirrored in it's graph quantization.
It was, in fact, the puzzling lack of births and deaths in a class of
(formal) solutions to
Ashtekar's constraints found by Rovelli and Smolin in \cite{RSloops} that
initially motivated
me to rederive the constraints. In these solutions
(called the RS solutions from here on)
the graph representation of the state, $\Psi[\Gamma]$, is the characteristic
function
of the graph class (= equivalence class of graphs under 3-diffeos connected to
the identity) of a graph, $\cg$, consisting of one or more intersection free
loops
carrying spin $\frac{1}{2}$. In other words, if $K_{\cg,\frac{1}{2}}$ is the
graph class in question, the wave function is of the form
\begin{equation} \label{RSsoln}
\Psi_{K_{\cg,\frac{1}{2}}}[\Gamma] = \left\{ \begin{array}{ll}
				1 & \mbox{if $\Gamma\in K_{\cg,\frac{1}{2}}$} \\
				0 & \mbox{otherwise}
			  \end{array}
		  \right.
\end{equation}
In such a state the number of loops is fixed, so there is zero amplitude for
births and deaths of loops.
Note that Rovelli and Smolin's argument to the effect that
$\Psi_{K_{\cg,\frac{1}{2}}}$ solves the quantized constraints goes through
unchanged if the loops are allowed to carry arbitrary spin, $j$.

Now suppose that a quantum theory of gravity possesses the spin $j$ RS
solutions.
Taking the $\hbar \rightarrow 0, \ \ j \sim O(1/\hbar)$ limit shows that
$\tilE$ flux loops can only evolve by 3-diffeos. No births or deaths are
allowed in the classical theory.\footnote{Strictly speaking it is possible
for the classical action to have stationary points that have zero weight
in the Feynman path integral. In this way a process that is quantum
mechanically
forbidden can be formally allowed in the classical theory.}
The theory is thus certainly not the Plebanski theory. In fact it will be
argued in section \ref{covariance} that the theory is not even fully 4-diffeo
invariant.

It seems, therefore, that the new constraint (\ref{generalized_SD}) leads to
births and deaths in the graph representation.

I should emphasize again, however, that, though motivated by quantum
considerations,
this paper deals exclusively with classical theory.

The remainder of this paper is organized as follows:

\noindent
In section \ref{covariance} the invariance of Ashtekar's theory under
all infinitesimal 4-diffeos\footnote{%
Matschull has claimed \protect\cite{Matschull94} that Ashtekar's theory is {\em
not} invariant
under all infinitesimal 4-diffeos when $\tilE$ is degenerate, contradicting the
results of
section \ref{covariance}. However, he now agrees that that result of
\protect\cite{Matschull94} is wrong \protect\cite{Mat_priv}.}
, and its non-invariance under
some finite 4-diffeos is established. Furthermore, it is argued that Ashtekar's
theory is not a gauge fixed version of any 4-diffeo invariant local theory,
because it does not posses a `2-sphere solution', which describes
an $\tilE$ flux loop being born from the vacuum, $\tilE = 0$, and eventually
disappearing again.

\noindent
In section \ref{Plebanski_action} Plebanski's
action is derived from the familiar Hilbert-Palatini action and the field
equations in Plebanski's variables are found.

\noindent
In section \ref{three_plus_one} a canonical formulation of Plebanski's version
of GR
in terms of Ashtekar's variables, $A$ and $\tilE$, is derived from Plebanski's
action.
(\ref{Gauss_law}) and (\ref{generalized_SD}) are the constraints of this
formulation.

\noindent
In section \ref{three_plus_one2} a more elegant, but equivalent, canonical
theory, in which
the field $\phi^{ij}$
in (\ref{generalized_SD}) (which is, in fact, the left-handed Weyl curvature)
and a conjugate momentum, $\tilpi_{ij}$, are treated as canonical coordinates.

\noindent
Section \ref{spt_sphere} develops a `2-sphere solution' to
Plebanski's spacetime field equations. In this solution both the
self-dual curvature, $F^i_{\mu\nu}$, and the orthonormal basis of self-dual
2-forms, $\Sigma_{i\,\mu\nu}$, have support on a (thickened) 2-sphere in
spacetime.

\noindent
In section \ref{canon_sphere} it is shown that this 2-sphere solution
solves the canonical theory of section \ref{three_plus_one}. How
the new Hamiltonian of section \ref{three_plus_one} generates the birth in this
solution is explained in detail.



\section{4-diffeomorphisms in Ashtekar's canonical theory}\label{covariance}

Let's begin by very briefly summarizing Ashtekar's canonical theory in our
notation.\footnote{%
This notation differs from that of \cite{Ashtekar_Tate}
mainly in that $SO(3)$ tensors are used in place of the corresponding $SU(2)$
spinors, and that the
fields, which can in general be complex, are defined so that when they are real
the Euclidean theory is recovered.}
No proofs will be given since they can be found in e.g. \cite{Ashtekar_Tate},
and most statements will in fact be special cases of results of Section
\ref{three_plus_one}.

The canonical coordinates are the fields $A^i_a$ and $\tilE^a_i$, which live
on 3-space $\Sg$ and have Poisson bracket
\be	\label{fund_Poisson}
\{A^i_a(x), \tilE^b_j(y) \} = \dg^i_j\dg^b_a\dg^3(x,y).
\ee
The constraints are
\bearr
\caltilG_i & = & D_a\tilE^a_i = 0	\label{Gauss1}			\\
\caltilV_a & = & \frac{1}{2}\eg_{abc}\tilB^{i\,b}\tilE^a_i = 0
					\label{vector1}			\\
\caltilS & = & \frac{1}{4}\eg_{abc}\tilB^{i\,a}\tilE^b_j\tilE^c_k \eg_i{}^{jk}
	= 0,				\label{scalar1}
\eearr
and the Hamiltonian is a sum of these constraints:
\bearr
H_{Ash} & = & -\int_\Sg \Lam^i\caltilG_i + N^a\caltilV_a + \subtil{N}\caltilS\
d^3x \label{H_Ash1}\\
	& = & -\int_\Sg (\Lam^i - N^a A^i_a)\caltilG_i + N^a\tilDg_a +
\subtil{N}\caltilS\ d^3x		\\
	& = & - G_{[\Lam - N^a A_a]} - \Dg_{\vec{N}} - S_{\subtil{N}},
\eearr
where $\tilDg_a = \caltilV_a + A_a^i\caltilG_i$, and $G_\Lam = \int_\Sg
\Lam^i\caltilG_i\ d^3x$,
$\Dg_{\vec{N}} = \int_\Sg N^a\tilDg_a\ d^3x$, and
$S_{\subtil{N}} = \int_\Sg \subtil{N}\caltilS\ d^3x$.

This particular decomposition of the Hamiltonian into integrated constraints
has the advantage that $G_\Lam$ and $\Dg_{\vec{N}}$ have simple
interpretations:
$G_\Lam$ generates $SO(3)_0$\footnote{$SO(3)_0$ is the part of $SO(3)$
connected to
the identity.} gauge transformations, $\Dg_{\vec{N}}$ generates
3-diffeomorphisms. The constraint algebra is
\be
\begin{array}{lll}
\!\!\{G_{\Lam_1},G_{\Lam_2}\} = -G_{[\Lam_1,\Lam_2]}	& & \\
 & & \\
\!\!\{G_{\Lam},\Dg_{\vec{N}}\} = G_{\pounds_{\vec{N}} \Lam}	&
\!\!\!\!\{\Dg_{\vec{N}_1},\Dg_{\vec{N}_2}\} = -\Dg_{[\vec{N}_1,\vec{N}_2]} &
\\
 & & \\

\!\!\{G_\Lam, S_{\subtil{N}}\} = 0	&
\!\!\!\!\{\Dg_{\vec{N}}, S_{\subtil{N}}\} = -S_{\pounds_{\vec{N}} \subtil{N}}	&
\!\!\!\!\{S_{\subtil{N}_1}, S_{\subtil{N}_2}\}	= \Dg_{\vec{K}} - G_{K^a A_a},
\end{array}
\ee
with $K^a = K^a(\subtil{N}_1,\subtil{N}_2) = \tilE^{i\,a}\tilE^a_i\:
(\subtil{N}_1\di_b\subtil{N}_2 - \subtil{N}_2\di_b\subtil{N}_1)$,
$[\Lam_1,\Lam_2]^i = \Lam_1^j\Lam_2^k
\eg^i{}_{jk}$, and $[\vec{N}_1,\vec{N}_2]^a = N_1^b\di_b N_2^a -
N_2^b\di_b N_1^a$.

That completes the summary of Ashtekar's theory. Now to diffeomorphisms.

Because the constraints (\ref{Gauss1}), (\ref{vector1}) and (\ref{scalar1}) are
first class and complete, all gauge transformations of the
classical state $(A,\tilE)$ are generated by $G_\Lam$, $\Dg_{\vec{N}}$ and
$S_{\subtil{N}}$. Can a subset of these gauge transformations be interpreted
as the group, $\DiffM$, of 4-diffeos of spacetime, $M$, connected to the
identity?

The history of $X \equiv (A, \tilE, \Lam, \vec{N}, \subtil{N})$ generated by
the Hamiltonian can be thought of as a field configuration on the spacetime
$M = \{(\calP,t)|\calP\in\Sg, t\in \Real \}$, in which the fields are described
in terms of quantities that refer (like vector components refer to a basis)
to the `slicing' $\{\Sg_t\}$, consisting of the equal $t$ 3-surfaces, and the
`threading' ${\cg_\calP}$, consisting of the constant $\calP\in\Sg$ worldlines.

Suppose another slicing and threading $\{\Sg'_{t'}\}$, $\{\cg'_{\calP'}\}$ is
defined
by acting on the $\Sg_t$ and $\cg_\calP$ with a 4-diffeo. What shall we take to
be
the corresponding fields $X' = (A',\tilE',\Lam',\vec{N}',\subtil{N}')$? The
quantities
$X$ can, of course, be written as functions of the new `coordinates'
$(\calP',t')$, but they still refer to the old slicing and threading, and the
$X$ need
not {\em a priori} transform as scalars. The only {\em a priori} restriction on
the (1 to 1)
map $X \rightarrow X'$ that will be made here is that it be local: $X'$ at a
spacetime
point $p$ depends only on $X$ and the 4-diffeo within an infinitesimal
neighborhood of
$p$. (If we think of the 4-diffeos as active instead of passive then
$X'$ should depend only on $X$ and the diffeo near the pre-image of $p$).

The question is now: can this `local' representation of diffeomorphisms carried
by the fields $X$ be chosen so that each is a  gauge transformation. In other
words,
can one define the map $X \rightarrow X'$ in such a way that the history
$X'(\calP',t')$
is a gauge transform of the history $X(\calP,t)$.

It turns out that the transformation of the canonical coordinates $(A,\tilE)$
generated by
\be	\label{four_diff_generator}
J_\xi = \int_\Sg \xi^0\tilde{\cal H} - \xi^a\tilDg_a d^3x
\ee
can be interpreted as a 4-diffeo by the infinitesimal vector field $\xi^\m$.
($\tilde{\cal H} = -(\Lam^i - N^a A^i_a)\caltilG_i - N^a\tilDg_a -
\subtil{N}\caltilS$ is Ashtekar's Hamiltonian density). One can see at once
that $J_\xi$ generates the correct transformation in two simple cases.
When $\xi^0$ is constant in 3-space and $\xi^a = 0$ $J_\xi$ generates
a time reparametrization $t \rightarrow t - \xi^0$. When $\xi^0 = 0$
$J_\xi$ generates 3-diffeos by $\vec{\xi}$.

The Lagrange multipliers are also transformed in a gauge transformation.
A gauge transformed history is, after all, a history generated from
the same initial data, but with altered Lagrange multipliers put into the
Hamiltonian. The transformation law of the Lagrange multipliers can be derived
from the requirement that the gauge transformed Hamiltonian generates the
gauge transformed history of the canonical coordinates. Mathematically this
requires
\be
\dg^o_\xi H = \frac{\di^o}{\di t} J_\xi + \{J_\xi, H\},
\ee
where the ${}^o$ means that the canonical coordinates are held fixed
and only the Lagrange multipliers vary.

A rather intricate, but conceptually straightforward, calculation yields
\bearr
\dg_\xi \subtil{N} & = & \di_0 [\xi^0\subtil{N}] +
\pounds_{\vec{\xi}}\subtil{N} -
2\subtil{N} N^a\di_a \xi^0											\label{delta_lapse}\\
\dg_\xi N^a & = & \di_0 [\xi^0 N^a + \xi^a] + \pounds_{\vec{\xi}} N^a
- N^a N^b \di_b \xi^0 - K^a (\xi^0\subtil{N},\subtil{N})
\label{delta_shift}\\
\dg_\xi \Lam^i & = & [\pounds_\xi A]^i_0,
\eearr
where I have defined $A^i_0 = \Lam^i$, $\pounds_{\vec{\xi}}$ and $\pounds_\xi$
are the three and four dimensional
Lie derivatives respectively, and $K^a(\subtil{N}_1,\subtil{N}_2)$ is defined
as in the constraint
algebra.

With these transformations, and
\be
\dg_\xi A^i_a = \{A^i_a, J_\xi\}\ \ \ \ \dg_\xi\tilE^a_i = \{\tilE^i_a, J_\xi\}
\ee
the objects
\be
A^i_\m = \left[ \begin{array}{c} \Lam^i \\ A^i_a \end{array}\right]
\ee
and $\Sg_{i\,\m\n}$, with
\bearr
\Sg_{i\,ab} & = & \frac{1}{2} \eg_{abc}\tilE^c_i \\
\Sg_{i\,0a} & = & \frac{1}{4}\subtil{N}\eg_{abc}\eg_i{}^{jk}\tilE^b_j\tilE^c_k
+ \frac{1}{2}\eg_{abc}\tilE^b_i N^c,
\eearr
transform as spacetime tensor fields when $A$ and $\tilE$ are solutions to the
evolution equations. That is
\bearr
\dg_\xi A^i_\m & = & \pounds_\xi A^i_\m 				\label{A_transformation}\\
\dg_\xi \Sg_{i\,\m\n} & = & \pounds_\xi \Sg_{i\,\m\n}.
\label{Sig_transformation}
\eearr
Again the calculation is conceptually straightforward but quite tedious.

When $rank\ \tilE \geq 2$ $A^i_a$, $\tilE^a_i$, $\Lam^i$, $N^a$, and
$\subtil{N}$ at a spacetime point $p$ are functions of $A^i_\m$ and
$\Sg_{i\,\m\n}$
at $p$, so the transformation of these fields generated by $\dg_\xi$ is also a
`local' representation of the corresponding diffeomorphism. When $rank\ \tilE <
2$
$\subtil{N}$, and possibly
$N^a$, are undetermined by $A^i$ and $\Sg_i$. However, the transformation of
these fields,
as determined by (\ref{delta_lapse}) and (\ref{delta_shift}), is still a local
representation
of the diffeo.\footnote{\label{lapse_shift_trans}%
{\em Note added in proof}: The transformation
$(\subtil{N}, N^a) \rightarrow (\subtil{N}', N'^a)$ corresponding to the
finite coordinate transformation $x^\m \rightarrow z^\ag$ is given by:
\bearr
\subtil{N}' & = & \frac{1}{\Omega}\bar{\subtil{N}} 						\\
N'^a & = & \frac{1}{\Omega}[\bar{N}^a(1 + \bar{N}^b\frac{\di x^0}{\di z^b})
+ \bar{\subtil{N}}^2 \bar{\tilE}^a_i\bar{\tilE}^{i\,b}\frac{\di x^0}{\di z^b}]
- \frac{\di z^a}{\di x^0}/\frac{\di z^0}{\di x^0},
\eearr
where the $\bar{\ }$ quantities result from the purely spatial coordinate
transformation $x^u \rightarrow z^a$ induced by the spacetime coordinate
transformation on each $z^0 = \mbox{\em constant}$ hypersurface:
\be
N^u = \frac{\di x^u}{\di z^a}\bar{N}^a \ \ \ \tilE^u_i = det[\frac{\di x^v}{\di
z^b}]^{-1}
\frac{\di x^u}{\di z^a}\bar{\tilE}^a_i\ \ \ \subtil{N}
= det[\frac{\di x^v}{\di z^b}]\bar{\subtil{N}},
\ee
and
\be
\Omega = \frac{\di z^0}{\di x^0}[(1 + \bar{N}^a\frac{\di x^0}{\di z^a})^2
+ \bar{\subtil{N}}^2 \bar{\tilE}^a_i\bar{\tilE}^{i\,b}
\frac{\di x^0}{\di z^a}\frac{\di x^0}{\di z^b}].
\ee}
Ashtekar's theory is, therefore, invariant under infinitesimal 4-diffeos.

Note also that $G_\Lam$ and $J_\xi$ span the algebra of gauge generators (=
first class
constraints).

What about finite 4-diffeos? On certain solutions with degenerate $\tilE$
the representation $\dg_\xi$
of the 4-diffeo generators cannot be integrated to give the whole proper
4-diffeo group $\DiffM$ (Because $\subtil{N}$ and $N^a$ blow up when some
generators are iterated, see footnote \ref{lapse_shift_trans}).

This is most easily seen in solutions in which $\tilE^a_i$ is a single
unknotted flux loop
and $A^i_a = 0$. Then
\be
\tilE^a_i = e_i\int_\cg \dg^3(x,z)dz^a,
\ee
where $e_i$ is constant and $\cg$ is diffeomorphic to a circle. In the gauge
$\Lam^i = 0$ the evolution of the fields is given by
\bearr
\dot{\tilE}^a_i & = & \{\tilE^a_i, H_{Ash}\} = \pounds_{\vec{N}} \tilE^a_i
\label{E_evol}\\
\dot{A}^i_a & = & 0		\label{A_evol}
\eearr
so $\tilE^a_i$ simply evolves by 3-diffeos.

In spacetime this solution is described by\footnote{%
$dz^{[\rho}dz^{\sg]} =
\frac{dz^{[\rho}}{d\sg^1}\frac{dz^{\sg]}}{d\sg^2}\:d^2\sg$ where
$\sg^1,\:\sg^2$ are right handed coordinates on the 2-surface.}
\bearr
\Sg_{i\,\m\n} & = & \frac{1}{2} e_i\eg_{\m\n\rho\sg}\int_C \dg^4(x - z)
dz^{[\rho}dz^{\sg]}			\label{cylinder_Sg}\\
A^i_\m & = & 0.		\label{cylinder_A}
\eearr

\begin{figure}
\centerline{\psfig{figure=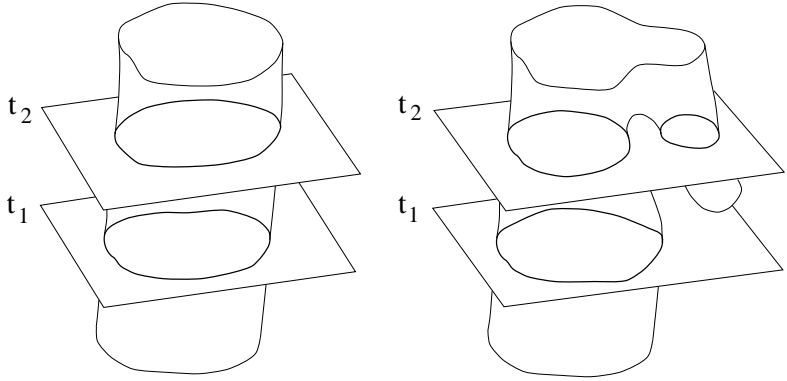,height=5cm}}
\caption{Two 4-diffeo equivalent evolutions of a flux loop. The cross sections
of the 2-surfaces indicated by the dark lines are the flux loops at different
times.}
\label{diff_birth}
\end{figure}

$C$ is the worldsheet of $\cg$. Since $\cg$ evolves only by 3-diffeos $C$ is
topologically a 2-cylinder. Clearly there are 4-diffeos of $\Sg_{i\,\m\n}$,
and thus of $C$ (or, equivalently, of the slicing and threading) such that
in the image the intersection $C\cap \Sg_t$ is not a single loop for all
$t$, but sometimes consists of several loops. (See Fig. \ref{diff_birth}). In
other words, there are
4-diffeos of the history of $(A,\tilE)$ in which births and deaths of flux
loops occur. This is, of course, not allowed by the evolution equations
(\ref{E_evol}) and (\ref{A_evol}), so these 4-diffeo equivalent histories are
not solutions.
Ashtekar's theory is thus not fully 4-diffeo invariant.

Could Ashtekar's theory be seen as a partly gauge fixed formulation of a
4-diffeo invariant theory? Let's, for the sake of argument, suppose that it is,
then the solutions of the
invariant theory would consist of all 4-diffeos of the solutions to Ashtekar's
theory. If the invariant theory is local, in the sense that it imposes only
local field equations on the fields, then if a field configuration solves
these equations in a basis of open sets it is a solution.

This is sufficient to show that the invariant theory has a `2-sphere' solution
in which $\Sg_i$ is supported on a 2-sphere $S$ in spacetime:
\bearr
\Sg_{i\,\m\n} & = & \frac{1}{2} e_i \eg_{\m\n\rho\sg}\int_S \dg^4(x -
z)dz^{[\rho}dz^{\sg]}		\label{sphere_Sg0} \\
A^i_\m & = & 0,		\label{sphere_A0}
\eearr
with $e_i$ constant. Within a sufficiently small open set one can always
pick a slicing and threading so $\tilE$ is a flux line evolving by 3-diffeos
only (and $A = 0$, $\Lam = 0$). The canonical evolution equations
(\ref{E_evol})
and (\ref{A_evol}), and constraints (\ref{Gauss1}),(\ref{vector1}), and
(\ref{scalar1}) thus hold within this open set, implying that the spacetime
field equations also do.

The 2-sphere solution has births and deaths in any slicing, so it is not
the diffeomorphic image of any solution of Ashtekar's theory.
Ashtekar's theory is thus not a gauge fixed version of a local 4-diffeo
invariant theory, because the gauge (slicing) in which there are no births and
deaths
does not exist for some solutions of any local theory having among its
solutions
all 4-diffeos of the solutions to Ashtekar's theory.\footnote{
The generator $J_\xi$ of 4-diffeos was an ansatz. Could all of $\DiffM$
be embedded in the gauge group if we started with a different generator?
Since $J_\xi$ and $G_\Lam$ span the gauge algebra any new 4-diffeo generator
$J'_\xi$ must be a combination of these: $J'_\xi = J_{\xi'[\xi]} +
G_{\Lam[\xi]}$.
This $J'_\xi$ generates mappings, $\phi'^{*}$, that take $\Sg_i = 0, A^i = 0$
to
$\Sg_i = 0, A^i = 0$. A local representation of a diffeomorphism only knows
that
$\Sg_i \neq 0$ on the support of $\Sg_i$. In the context of solution
(\ref{cylinder_Sg}), (\ref{cylinder_A}) this means that in coordinate space
$\phi'^{*}\Sg_i$, the image of $\Sg_i$ under the mapping
$\phi'^{*}$ associated with the diffeomorphism $\phi$ in the representation
generated by $J'$, has support only on $\phi C$. Considering again
infinitesimal
transformations we see that this means that displacing $C$ by $\xi$ and by
$\xi'$
produces $\phi C$ and a subset of $\phi C$, respectively.
If we now consider $C$ which has been displaced
a little bit near a given point we see that locality implies that $\xi' = \xi$
wherever
$\phi'^{*}$ does not map $\Sg_i$ to zero. But $[\phi'^{*}\Sg_i](x-\xi') \neq 0$
when
$x \in C$, so, in fact, $\xi' = \xi$ on all of $C$. Since $C$ can be chosen to
go
 through any
point in spacetime this equality actually holds everywhere. Modulo $SO(3)_0$
gauge
transformations, the local representation of 4-diffeo's as gauge
transformations is unique!
Hence mappings of solutions to non-solutions,
like those found in the representation of $\DiffM$ generated by $J$, occur in
all such representations of $\DiffM$ - the theory is intrinsically not
invariant under
the full $\DiffM$ group.}

Of course, the truncation of the 4-diffeomorphism symmetry we have seen in
Ashtekar's theory also occurs in standard Lorentzian canonical GR, because
of the requirement that the $\Sg_t$ be spacelike Cauchy surfaces. This
condition
also excludes
some solutions of GR (solutions with closed timelike curves) from the
canonical theory. This is sometimes even seen as an advantage of the
canonical theory over the fully 4-diffeo invariant version because it
ensures causality.

Here we have not imposed the Lorentzian reality conditions. If all the fields
are taken to be real a Euclidean theory is obtained.
Nevertheless an extension of the notion of causality to degenerate Euclidean
geometries , such as the requirement that there be no births or deaths of
flux loops, might justify the non-invariance of Ashtekar's theory. Such a
causality requirement is not entirely unreasonable since births and deaths
are in fact `uncaused' (gauge) - they cannot be predicted form the canonical
initial data. Whether such causality conditions {\em should} be applied,
especially in the
quantum theory, is another question. The issue of causality in degenerate
geometries needs to be explored further.


\section{The Plebanski action} \label{Plebanski_action}

The Plebanski action will be used to define GR in this paper. In particular,
the canonical theory of Section \ref{three_plus_one} is derived from it. It is
classically equivalent to the Einstein-Hilbert (EH) action except in that,
because it is well defined on a larger class of `geometries' than the
EH action, the space of classical solutions it defines is larger than
that of the EH action. Not all extrema of the Plebanski action
correspond to invertible metrics $g_{\m\n}$.

In this section the definition of the Plebanski action, and its relation to the
EH action are reviewed (chiefly following \cite{CDJ} and \cite{Ashtekar_Tate}),
and the field equations defined by the Plebanski action are given.

Let's begin by reviewing the concept of self-duality, taking the opportunity
to fix notation along the way. In this paper we are concerned with
(complex) Euclidean GR.
The internal symmetry group is thus $SO(4)$, that is, gauge transformations
of the vierbein $e_I{}^\m$ preserve the internal metric $\dg_{IJ}$.
($SO(4)$ indices, which range over $\{0,1,2,3\}$, are represented by
upper case latin letters from the middle of the alphabet: $I$, $J$, $K$, ... .
Spacetime indices are represented by lower case Greek letters.)

$SO(4)_0$\footnote{$SO(4)_0$ is the part of $SO(4)$ that is connected to the
identity.}
is the direct product of two factors of $SO(3)_0$, which will be called
$SO(3)_L$ and $SO(3)_R$: $SO(4)_0 = SO(3)_L \otimes SO(3)_R$. As a result
$SO(4)$ tensors in the adjoint representation, and thus $SO(4)$ connections
and curvatures, can be split into ``self-dual" and ``anti-self-dual"
components,
which transform under $SO(3)_L$ and $SO(3)_R$ respectively.\footnote{%
In a spinor (double-valued) representation of $SO(3)_0$ self-dual tensors are
left-handed spinors
and anti-self-dual tensors are right-handed spinors, hence the subscripts
$L$ and $R$ on the self-dual and anti-self-dual $SO(3)_0$ factors in
$SO(4)_0$.}
Let's see how this comes about.

The $SO(4)$ dual of an antisymmetric tensor $a^{IJ}$ is defined as
\be
a^{\star\,IJ} = \frac{1}{2}\eg^{IJ}{}_{KL}a^{KL}.
\ee
$SO(4)_0$ transformations leave the duality operator
$\frac{1}{2}\eg^{IJ}{}_{KL} = \frac{1}{2} \epsilon^{IJMN}\dg_{MK}\dg_{NL}$
invariant. Thus the adjoint representation, which acts on
antisymmetric tensors $a^{[IJ]}$, reduces to a sum of representations acting
in the two eigensubspaces of the duality operator, namely the self-dual
representation acting on self-dual tensors $a^{+} = a^{+\star}$, and an
anti-self-dual rep. acting on anti-self-dual
tensors $a^{-}= - a^{-\star}$.
Note that any antisymmetric tensor $a^{IJ}$ can be split into a self-dual and
an anti-self-dual component according to
\begin{equation}
a^{\pm} = \frac{1}{2}[ a \pm a^\star ].
\end{equation}

(Anti-)self-dual tensors have only three independent components. According
to their definition $a^{\pm ij} = \pm \eg^{ij}{}_k a^{\pm 0k}\ $
($i,j,k\in\{1,2,3\}$) so we may take the independent components to be
\begin{equation} \label{SDcomponents}
a^{\pm i} = \pm 2 a^{\pm 0i}.
\end{equation}

The $so(4)$ generators are themselves adjoint rep. tensors. Decomposing
these into their self-dual and anti-self-dual components lets us rewrite the
$so(4)$ commutation relations\footnote{%
In the respective fundamental representations $[G_{IJ}]_M{}^N = -\dg_{M[I}
\dg_{J]}{}^N$ and $[G^\pm_i]_m{}^n = \eg_{mi}{}^n$.}
\begin{equation}
[G_{IJ},G_{KL}] = -\{ \dg_{I[K} G_{L]J} - \dg_{J[K} G_{L]I} \}
\end{equation}
as
\begin{eqnarray}
[G^+_i, G^+_j] & = & \eg_{ij}{}^k G^+_k \\
{[}G^+_i, G^-_j] & = & 0 \\
{[}G^-_i, G^-_j] & = & \eg_{ij}{}^k G^-_k,
\end{eqnarray}
which defines two commuting $so(3)$ algebras. In other words
$SO(4)_0 = SO(3)_L \otimes SO(3)_R$ where, in the adjoint rep. of $SO(4)$
$SO(3)_L$
acts on self-dual tensors, and $SO(3)_R$ acts on anti-self-dual tensors.
Specifically, $a^{+i}$ and $a^{-i}$ transform as, respectively,
$SO(3)_L$ and $SO(3)_R$ vectors.\footnote{%
The action of the generators on adjoint rep. tensors can be represented
as the commutator of the tensors with the corresponding fundamental rep.
generators. Since the generators of $SO(3)_L$ commute with anti-self-dual
tensors, which are linear combinations of the generators of $SO(3)_R$ in the
fundamental rep of $SO(4)$, $SO(3)_L$ acts only on self-dual tensors.
Specifically, $(G^+_{i\,\adj}\,a^+)^j\,G^+_{j\,\fun} = [G^+_{i\,\fun},
a^{+j}G^+_{j\,\fun}] =
a^{+n}\eg_{in}{}^j G^+_{j\,\fun}$, so $(G^+_{i\,\adj}\,a^+)^j =
\eg^j{}_{in}a^{+n}$. $SO(3)_R$
acts similarly on anti-self-dual tensors.}
generators, $SO(3)_L$ From here on $SO(3)$ indices, which run
over $\{1,2,3\}$, will always be represented by lower case latin letters,
$i$,$j$,$k$,..., from the middle of the alphabet. Note that upper and lower
$SO(3)$ indices are equivalent since the $SO(3)$ metric is $\dg_{ij}$.

It is a remarkable fact that an action can be written for GR involving only
self-dual, or left-handed, quantities, so that the internal symmetry group
becomes simply $SO(3)$. The Plebanski action is such an action. It is
\begin{equation} \label{Pleb_a}
I = \int \frac{1}{2} \Sigma_i \wedge F^i
	- \frac{1}{4} \phi^{ij} \Sigma_i \wedge \Sigma_j.
\end{equation}
$F^i_{\m\n} = 2\di_{[\m}A^i_{\n]} + \eg^i{}_{jk}A^j_\m A^k_\n$
is the curvature of the $SO(3)$ connection $A^i_\m$.
$\Sg_{i\,\m\n}$ is an $SO(3)$ vector 2-form, and $\phi^{ij}$, which is required
to be trace free ($\phi^{ij}\dg_{ij} = 0$) acts as a Lagrange multiplier.

The field equations implied by the stationary of $I$ under variations
of $\phi$, $A$ and $\Sigma$ are, respectively
\bearr
\Sg_i\wedge\Sg_j & \propto & \delta_{ij}\eg  \label{fieldeq1} \\
D\wedge\Sg_i & = & 0			\label{fieldeq2} \\
F^i - \phi^{ij}\Sg_j & = & 0.		\label{fieldeq3}
\eearr
Here $\eg_{\kappa\lam\m\n}$ is the spacetime antisymmetric symbol, which can be
thought of as
the coordinate volume form. On tensors with only $SO(3)$ indices $D$ is the
covariant derivative with connection $A$. For example,
$D_\m v^i = \di_\m v^i + A^j_\m \eg^i{}_{jk} v^k$ ($\eg^i{}_{jk}$ is the $j$th
generator of $SO(3)$ in the fundamental rep.). $D_{[\m}\Sg_{\n\lam]}$ in
(\ref{fieldeq2})
is evaluated using a torsionless extension of $D$ to spacetime tensors. Which
torsionless extension is used is immaterial because of the antisymmetrization
of the spacetime indices.

In appendix \ref{sig_tet} it is shown that if $v = \frac{2}{3}\Sg_i\wedge\Sg^i
\neq 0$ then (\ref{fieldeq1})
implies that there exists a non-singular tetrad $e^I{}_\m$, unique up to
$SO(3)_R < SO(4)$ transformations on the internal index $I$, such that
\be
\Sg_i = \frac{1}{2} e^0\wedge e^i + \frac{1}{4}\eg_{ijk} e^j\wedge e^k.
\label{Sgtetform}
\ee
In other words, $\Sg_i$ is the self dual part of $\frac{1}{2}e^I\wedge e^J$
with respect to the internal 4-metric $\dg_{IJ}$. Note that $e^I{}_\m$ forms an
orthonormal
tetrad with respect to the spacetime metric $g_{\m\n} = e^I{}_\m e^J{}_\n
\dg_{IJ}$, and that
$v = e^0\wedge e^1\wedge e^2 \wedge e^3$ is the volume form of this metric.

In the following it will be shown that if $v \neq 0$ in an open region, $U$,
then the field
equations (\ref{fieldeq2}) and (\ref{fieldeq3}) imply that the metric $g_{\m\n}
= e^I{}_\m e^J{}_\n \dg_{IJ}$ solves Einstein's vacuum field equation,
$R_{\m\n}[g] = 0$, in $U$, and $A^i_\m$ is the self-dual part of the metric
compatible $SO(4)$
connection $\omega_\m^{IJ} = e_\ag{}^J [\di_\m e^{\ag I} +
\{{}^{\,\ag}_{\bg\m}\} e^{\bg I}]$. ($\{{}^{\,\ag}_{\bg\m}\}$ is the spacetime
connection of $g$).

Conversely, taking as $(\Sg_i, A^i )$ the self-dual parts of $(\frac{1}{2}
e^I\wedge e^J,
\ \omega^{IJ})$ in a solution to Einstein's field equation on $U$ yields a
solution to
(\ref{fieldeq1}), (\ref{fieldeq2}) and (\ref{fieldeq3}) (with suitably chosen
$\phi$)\footnote{
What is the value of $\phi$ on a solution of Einstein's equation? The curvature
2-form can be expanded as
\be
F^i_{\m\n} = F^i_{IJ} e^I{}_{[\m} e^J{}_{\n]} = F^{++\,ij} \frac{1}{2}[e\wedge
e]^{+}_{j\,\m\n}
			+ F^{+-\,ij} \frac{1}{2}[e\wedge e]^{-}_{j\,\m\n},
\ee
where in the last expression $F^i_{IJ}$ and $\frac{1}{2} e^I\wedge e^J$ have
been expanded
into self-dual and anti-self-dual components with respect to the indices $I\
J$. On
solutions $\Sg_j = \frac{1}{2}[e\wedge e]^+_j$ so field equation
(\ref{fieldeq3}) shows that, firstly, $F^i_{IJ}$
is self-dual on $I\ J$ and, secondly, $F^{ij} = F^{++ij} = \phi^{ij}$.
On vacuum solutions $F$
is the self-dual part of the Riemann curvature, which, in turn, equals the Weyl
curvature.  The
$\phi^{ij}$ are therefore the internal components (components in the basis
$\Sg_i$) of the self-dual Weyl curvature. $\phi^{ij}$ is equivalent to the
Weyl curvature spinor. Explicitly this spinor is
$\Psi_{ABCD} = \phi^{ij} \sg_{i\,AB} \sg_{j\,CD}$, where the $\sg_i$ are the
Pauli spin matrices. }
in which $v\neq 0$ in $U$. The set of solutions to (\ref{fieldeq1}),
(\ref{fieldeq2}) and
(\ref{fieldeq3}) with $v\neq 0$ is thus just the set of solutions to Einstein's
vacuum
equation.

There are also solutions to (\ref{fieldeq1}), (\ref{fieldeq2}) and
(\ref{fieldeq3})
with $v = 0$. These do not correspond to solutions of Einstein's equations in
good coordinates,
since the geometrical volume of finite coordinate volumes is zero,\footnote{%
``Good coordinates'' are diffeomorphic to normal coordinates. This requires
the Jacobian of the transformation to normal coordinates, which is
$[\eg^{\m\n\sg\rho}v_{\m\n\sg\rho}]^{-1}$, to be everywhere finite.}
 and some do not correspond
to any coordinatization of a solution to Einstein's equations. Such solutions
will be the focus of this paper.

Now let's prove the equivalence of the $v \neq 0$ sector of Plebanski's theory
with standard GR.
We begin by restricting the Plebanski action to solutions of (\ref{fieldeq1}).
The $\Sg_i$
are then parametrized by $e^I{}_\m$ according to (\ref{Sgtetform}).
Specifically, $\Sg_i$ is
the self-dual part of $\frac{1}{2}e^I\wedge e^J$ with respect to the internal
metric
$\dg_{IJ}$. Thus on solutions of (\ref{fieldeq1})
\be
I = \frac{1}{2}\int \Sg_i\wedge F^i = \frac{1}{4}\int e_I \wedge e_J \wedge
F^{IJ},
\ee
where $F^{IJ}$, defined by $F^{0i} = \frac{1}{2}F^i$, $F^{ij} =
\frac{1}{2}\eg^{ij}{}_k F^k$,
is the curvature of the self-dual connection $A^{IJ}$, defined similarly by
$A^{0i}
= \frac{1}{2}A^i$, $A^{ij} = \frac{1}{2}\eg^{ij}{}_k A^k$.

Because $F^{IJ}$ is self-dual
\bearr
I = \frac{1}{8}\int e^I\wedge e^J\wedge F^{KL} \eg_{IJKL} & = &
\frac{1}{8}\int \eg^{\kappa\lam\m\n} e^I{}_\kappa e^J{}_\lam F^{KL}_{\m\n}
\eg_{IJKL}\ d^4x \\
	& = & \frac{1}{2} \int e\ e_K{}^\m e_L{}^\n F^{KL}_{\m\n}\ d^4x,
\label{SJS_action}
\eearr
where $e = det[ e^I{}_\m]$. (\ref{SJS_action}) is the self-dual action for GR
found by Samuel
\cite{Samuel87} and Jacobson and Smolin \cite{Jacobson_Smolin87},
\cite{Jacobson_Smolin88a}.

Let $\nabla$ be the (unique) torsionless derivative compatible with the
spacetime metric
$g_{\m\n} = e^I{}_\m e^J{}_\n \dg_{IJ}$, and extend its action to internal
indices by
requiring $\nabla e^I{}_\m = 0$. The internal connection coefficients of
$\nabla$ are then
$\omega_\m^{IJ} = e_\ag{}^J [\di_\m e^{\ag I} +
\{{}^{\,\ag}_{\bg\m}\} e^{\bg I}]$.
Define $C^+$ as the difference between the self-dual connection $A^{IJ}$ and
the self-dual
part $\omega^{+IJ}$ of the metric connection: $C^{+IJ}_\m = A^{IJ}_\m -
\omega^{+IJ}_\m$.
$F$ can then be expanded as a sum of the curvature, $R^+$, of $\omega^+$, and
terms in $C^+$:
\be
F^{IJ}_{\m\n} = R^{+IJ}_{\m\n} + 2\nabla_{[\m} C^{+IJ}_{\n]} + 2 C^{+IM}_{[\m}
C^+_{\n]M}{}^J
\ee
$R^+$ is the self-dual part of the Riemann curvature tensor.
Thus, from (\ref{SJS_action}),
\be
I = \frac{1}{2} \int e\ e_K{}^\m e_L{}^\n \ \{R^{+KL}_{\m\n} + 2\nabla_\m
C^{+KL}_\n
+ 2 C^{+KM}_{[\m} C^+_{\n]M}{}^L \}\ d^4x. \label{cplus1}
\ee
Consider the first term in (\ref{cplus1}).
\bearr
e\ e^\m_K e^\n_L\ R^{+KL}_{\m\n} & = & \frac{1}{2}\, e\ [e^\m_K e^\n_L
R^{KL}_{\m\n} +
\frac{1}{2}e^\m_K e^\n_L \eg^{KL}{}_{IJ}R^{IJ}_{\m\n}] \\
	& = & \frac{1}{2}\sqrt{g}[R +
\frac{1}{2}\eg^{\kappa\lam\m\n}R_{\kappa\lam\m\n}]
= \frac{1}{2}\sqrt{g}R,
\eearr
since $R^\kappa{}_{[\lam\m\n]} = 0$. The first term in (\ref{cplus1}) is thus
just the Einstein-Hilbert
action $I_{EH} = \frac{1}{4}\int R\sqrt{g} d^4x$. The second term in the
integrand of
(\ref{cplus1}) is a divergence, since $\nabla  e_I{}^\m = 0$. $I$ is thus a sum
of the
Einstein-Hilbert action, a surface term, and a potential term quadratic in the
field $C^+$ which
does not enter the Einstein-Hilbert action.
\be
I = I_{EH} + \mbox{\em surface term} + \int e\ e^\m_{[K} e^\n_{L]}\,
C^{+KM}_{\m}
C^+_{n\, M}{}^L\ d^4x.
\label{cplus2}
\ee

By splitting $C^+$ into judiciously chosen components the $C^+$ potential term
can be diagonalized.
Define $C^{+JK}_I = e_I{}^\m C^{+JK}_\m$, then let $C^{+K} = C^{+IK}_I$ and
$\bar{C}^{+IJK} = C^{+IJK} - \frac{2}{3}\dg^{I[J}C^{+K]} - C^{+[IJK]}$ (so that
$\dg_{IJ}\bar{C}^{+IJK} = 0$ and $\bar{C}^{+[IJK]} = 0$). The three tensors,
$C^{+K}$, $C^{+[IJK]}$,
and $\bar{C}^{+IJK}$ are independent components of $C^{+IJK}$, with no
constraints correlating
them. In terms of these components the $C^+$ potential term is
\be
V = \int e\ \{ -\frac{1}{3} C^{+M}C^+_M +
\frac{1}{4}\bar{C}^{+KLM}\bar{C}^+{}_{KLM} - \frac{1}{2}
C^{+[KLM]}C^+{}_{[KLM]} \}\ d^4x	\label{diag_potential}
\ee
This potential clearly has no zero modes, so extremizing $I$ with respect to
$C^+$ (equivalently,
solving (\ref{fieldeq2}), $\frac{\dg I}{\dg A} = 0$) requires $C^+ = 0$, in
other words,
$A = \omega^+$. Furthermore, on the extremum with respect to $C^+$, $I$ is
equivalent
to $I_{EH}$. That is to say, the only remaining field equation,
(\ref{fieldeq3}),
$\frac{\dg I}{\dg\Sg} = 0$, becomes $0 = \frac{\dg I}{\dg g_{\m\n}}
= \frac{\dg I_{EH}}{\dg g_{\m\n}}$.\footnote{
It is assumed that the functional derivative is taken in the interior of the
spacetime volume
of the variational problem so that the surface term in $I$ does not
contribute.}
As is well known, $\frac{\dg I_{EH}}{\dg g_{\m\n}} = 0 \Leftrightarrow R_{\m\n}
+ \frac{1}{2}R g_{\m\n}=0
\Leftrightarrow R_{\m\n} = 0$, which is just the Einstein vacuum field
equation. This proves that
solutions of (\ref{fieldeq1}), (\ref{fieldeq2}), and (\ref{fieldeq3}) with
$v = \frac{2}{3}\Sg_i\wedge\Sg^i \neq 0$ correspond to solutions of Einstein's
equation. The converse is clear.

\section{Canonical formulation in terms of Ashtekar's variables}
\label{three_plus_one}

To derive the canonical theory corresponding to the Plebanski action we
begin by choosing a slicing of spacetime $M$ into 3-surfaces $\Sg_t$,
parametrized by `time' $t\in \Real$, and all diffeomorphic to one another
(and thus to $\Sg = \Sg_0$). The $\Sg_t$ will not be assumed to be
`spacelike', i.e. to have a positive definite, and thus non-degenerate,
spatial metric, since in many of the degenerate solutions we are interested
in this condition is not met by any slicing. Since this paper is not concerned
with the effects of a non-trivial topology of $\Sg$ or $M$ these are assumed,
for simplicity and definiteness, to be diffeomorphic to $S^3$ and $S^3\times
\Real$ respectively. The canonical variables will be fields living on
$\Sg$.

\begin{figure}
\centerline{\psfig{figure=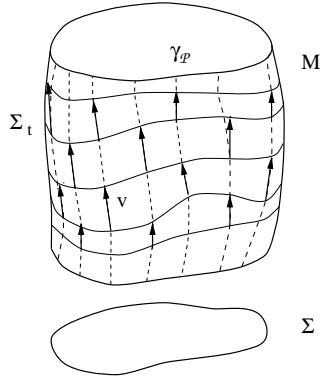,height=5cm}}
\caption{Schematic illustration of spacetime $M$ and space $\Sg$, in which
$M = S^3\times\Real$ and $\Sg = S^3$ are represented by $S^1\times\Real$
and $S^1$ respectively. The slicing, $\{\Sg_t|t\in\Real\}$, and threading,
$\{\cg_\calP|\calP\in \Sg\}$, of $M$ are indicated, as well as the `time flow'
vector field $v = \frac{d}{dt}|_{\cg_\calP}$.}
\label{slice_thread}
\end{figure}

In addition to the slicing we also need to choose a `threading', a congruence
of curves, $\{\cg_{\cal P} | {\cal P}\in \Sg\}$, transverse to the $\Sg_t$ and
filling spacetime, which mark the world lines of `the same point' in
3-space $\Sg$. The solutions to the canonical theory will correspond to
the evolution, in $t$, of the fields on the slices $\Sg_t$ in a solution to the
spacetime field equations, with the time derivative at a point
${\cal P} \in \Sg$ in the canonical theory corresponding to $d/dt$ along
$\cg_{\cal P}$ in spacetime. $M$, $\Sg$ and the slicing and threading are
illustrated schematically in Fig. \ref{slice_thread}.

The tangent vector field, $v = \frac{d}{dt}|_{\cg_{\cal P}}$ of the
$\cg_{\cal P}$ is called the ``time flow vector". In the standard treatment
of canonical GR the metric is used to decompose $v$ into a piece $N^\m$ tangent
to $\Sg_t$ and a piece $N n^\m$ normal to $\Sg_t$, where $n$ is the unit
future pointing normal to $\Sg_t$. $N$ is called the lapse, and $N^\m$ is
called
the shift. This decomposition is not always well defined in the degenerate
solutions we are considering, so it will not be made here.

Once a slicing and a threading has been chosen $v$ and $dt$ can be used to make
a $3+1$ decomposition of the tensor fields appearing in the action.
That is, each such tensor is decomposed into spatial ($\Sg_t$) tensor
components. In local coordinates, $x^\m$, adapted to the slicing and threading
in that $x^0 = t$ and $\frac{dx^a}{dt}|_{\cg_{\cal P}} = 0$, this boils
down to writing the Lagrangian density as a sum of terms in which each
spacetime index is replaced by $0$ or a spatial index.
The Plebanski action (\ref{Pleb_a}) becomes
\be     \label{split_pleb}
I = \int \eg^{0abc}\Sg_{i\,bc} F^i_{0a} + \eg^{0abc} \Sg_{i\,0a} F^i_{bc}
	- \phi^{ij}\eg^{0abc}\Sg_{i\,0a}\Sg_{j\,bc}\ d^4x.
\ee
Here $\eg$ is the antisymmetric symbol with $\eg^{0123} = 1$.

(\ref{split_pleb}) can be put in a nice form using the definitions
$\eg^{abc} = \eg^{0abc}$, $\tilde{B}^{i\,a}
= \eg^{abc} F^i_{bc}$ and
\be   \label{defE}
\tilde{E}^a_i = \eg^{abc}\Sg_{i\,bc},
\ee
and the identity $F^i_{0a} = \di_0 A^i_a - D_a A^i_0$ (in which $A^i_0$ is
differentiated as though it were an $SO(3)$ vector). After an integration
by parts (\ref{split_pleb}) becomes
\bearr
I & = & \int \ \int_{\Sg_t} \tilE^a_i\,\di_0 A^i_a + A^i_0 D_a \tilE^a_i
	+ \Sg_{i\,0a}\tilB^{i\,a} -  \Sg_{i\,0a}\phi^{ij}\tilE^a_j\ d^3x\ dt
\label{split_pleb2}  \\
  & = & \int \ \int_{\Sg_t} \tilE^a_i\di_0 A^i_a  - \tilde{\cal H}\ d^3x\ dt.
\label{split_pleb3}
\eearr
Recall that $\Sg_t$ is assumed closed, so there are no boundary terms.

Plebanski's action can thus be seen as a phase space action for GR. One can
read off that $\tilE^a_i(x)$ is the momentum conjugate to $A^i_a(x)$, and that
the Hamiltonian density is
\be  \label{Hdensity}
\tilde{\cal H} = - A^i_0 D_a \tilE^a_i
	   - \Sg_{i\,0a}[\tilB^{i\,a} -  \phi^{ij}\tilE^a_j].
\ee

$A^i_0$, $\Sg_{i\,0a}$, and $\phi^{ij}$ enter (\ref{split_pleb2}) as Lagrange
multipliers. The classical state, $(A,\tilE)$, must therefore satisfy the
constraints
\setcounter{eqnum_mem}{\value{equation}}
\setcounter{equation}{\value{Gaussnum}}
\bearr
D_a \tilde{E}^a_i & = & 0  \\
\setcounter{equation}{\value{gSDnum}}
\exists \phi^{ij} \mbox{trace free, symmetric} & \suchthat & \tilde{B}^{i\,a} -
\phi^{ij} \tilde{E}^a_j = 0.
\eearr
\setcounter{equation}{\value{eqnum_mem}}

These constraints are the spatial parts of the field equations (\ref{fieldeq2})
and (\ref{fieldeq3}):
\be
\begin{array}{rclcl}
D\wedge\Sg_i & = & 0 	& \implies & 0 = D_{[a}\Sg_{i\, bc]} = \frac{1}{3!}
D_d \tilE^d_i \eg_{abc}	\\
F^i - \phi^{ij}\Sg_j & = & 0	& \implies & 0 = F^i_{ab} - \phi^{ij}
\Sg_{j\,ab} = \frac{1}{2}\eg_{abc} [\tilB^{i\,a} - \phi^{ij}\tilE^a_j ]
\end{array}
\ee
The time components of these equations give the evolution of $A$ and $\tilE$,
in terms of $A^i_0$, $\Sg_{i\,0a}$, and $\phi^{ij}$.

Stationarity of the action with respect to variations of $\phi^{ij}$
implies field equation (\ref{fieldeq1})
\be   \label{Sigma_constraint}
\begin{array}{rclcl}
\Sg_i \wedge \Sg_j & \propto & \dg_{ij} & \Leftrightarrow &
\Sg_{(i\,0a}\tilE^a_{j)}\propto \dg_{ij},
\end{array}
\ee
which places no constraint on the state, $(A,\tilE)$, since $\Sg_{i\,0a}$ is a
Lagrange multiplier and
may thus be freely chosen. However, for a given state it constrains
$\Sg_{i\,0a}$, which
restricts the possible evolutions of the state.

(\ref{Gauss_law}) and (\ref{generalized_SD}) are the primary constraints.
In fact, they are the complete constraints, since they are preserved by the
Hamiltonian evolution, without further conditions on the state.
(However, the conditions (\ref{Sigma_constraint}), (\ref{Sg_const2}), and
(\ref{Sg_const3})
on the Lagrange multiplier $\Sg_{i\,0a}$, are necessary).

Before proving the completeness of the constraints (\ref{Gauss_law}) and
(\ref{generalized_SD}), let's pause to understand what we have found so far.

(\ref{generalized_SD}) is not of the usual form
``{\em constraint function = 0}". Rather,
it demands merely that, for any admissible $(A, \tilE)$ there {\em exists}
a traceless, symmetric $\phi^{ij}$ such that
$\tilB^{i\,a} - \phi^{ij}\tilE^a_j$ vanishes.

The content of (\ref{generalized_SD}) becomes clearer when $\phi$ is
eliminated.
When $rank\tilE = 3$ ($\tilE^a_i$ invertible)
\be
\phi^{ij} = \tilB^{i\,a}\subtil{E}^{-1\,j}_a = \tilB^{i\,a}\eg^{jkl}\eg_{abc}
\tilE^b_k\tilE^c_l /2 det\tilE.
\ee
The constraints arise from the requirement that $\phi^{ij}$ be symmetric and
trace free.

Symmetry requires
\bearr
0 & = & \tilB^{[i\,a}\eg^{j]kl}\eg_{abc}\tilE^b_k\tilE^c_l \\
  & = & -\eg^{ijk}\eg_{abc}\tilB^{l\,a}\tilE^b_l\tilE^c_k,
\eearr
which holds if and only if
\be
\eg_{abc}\tilB^{i\,a}\tilE^b_i = 0.	\label{vector_1}
\ee
The tracelessness of $\phi^{ij}$ requires
\bearr
0 & = & \dg_{ij}\ \tilB^{i\,a}\eg^{jkl}\eg_{abc}\tilE^b_k\tilE^c_l\\
  & = & \eg^{ijk}\eg_{abc}\tilB^{i\,a}\tilE^b_j\tilE^c_k.  \label{scalar_1}
\eearr
(\ref{vector_1}) and (\ref{scalar_1}) (and (\ref{Gauss_law})) are just
Ashtekar's constraints. As shown in \cite{CDJM}, when $rank\tilE = 3$,
the Plebanski action leads exactly to Ashtekar's canonical theory. It
is less obvious, but nevertheless true, that (\ref{generalized_SD}) is
equivalent to Ashtekar's constraints (\ref{vector_1}) and (\ref{scalar_1})
also when $rank\tilE = 2$. This is shown in Appendix \ref{rank2app}.

When $rank\tilE = 1$. $\tilE^a_i$ is of the form $e_i\tilw^a$.
(\ref{generalized_SD}) requires that $\tilB^{i\,a}$ is also proportional
to $\tilw^a$: $\tilB^{i\,a} = b^i\tilw^a$, with $b^i - \phi^{ij}e_k = 0$.
A symmetric, traceless $\phi$ can always be found which satisfies this last
condition. Thus, when $rank\tilE = 1$ (\ref{generalized_SD}) is equivalent to
\be
\tilB^{i[a}\tilE^{b]}_j = 0.	\label{rank1_const}
\ee
When $\tilE^a_i = 0$ (\ref{generalized_SD}) is equivalent to
\be
\tilB^{i\,a} = 0. 		\label{rank0_const}
\ee

Summarizing:
\be     \label{content_SD}
\mbox{
\begin{tabular}{|c|lr|}   \hline
$rank\tilE$ & \multicolumn{2}{c|}{(\protect\ref{generalized_SD}) equivalent to}
\\ \hline
$3$ or $2$  	& $0 = \eg_{abc}\tilB^{i\,a}\tilE^b_i$ & (\protect\ref{vector_1})
\\
   & $0 = \eg^{ijk}\eg_{abc}\tilB^{i\,a}\tilE^b_j\tilE^c_k$ &
(\protect\ref{scalar_1}) \\
\hline
1 & $0 = \eg_{abc}\tilB^{i\,a}\tilE^b_j$	&  (\protect\ref{rank1_const}) \\
\hline
0 & $0 = \tilB^{i\,a}$		& (\protect\ref{rank0_const}) \\ \hline
\end{tabular}
}
\ee
Note that both (\ref{rank1_const}) and (\ref{rank0_const}) imply
(\ref{vector_1}) and (\ref{scalar_1}), so (\ref{Gauss_law}) and
(\ref{generalized_SD}) always imply Ashtekar's constraints. However, when
$rank\tilE\leq 1$, the converse is not true.
The solution set of (\ref{Gauss_law}) and (\ref{generalized_SD}) is the
Ashtekar
constraint surface with parts of the surface $rank\tilE\leq 1$ cut out.

\begin{figure}
\centerline{\psfig{figure=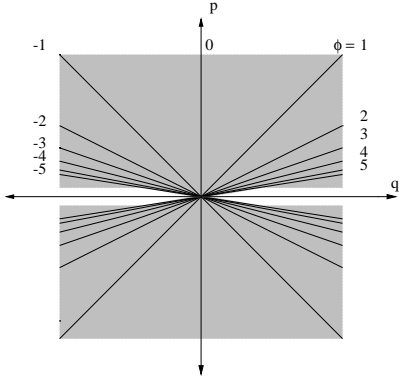,height=5cm}}
\caption{The solutions, $(q,p)$, to the constraint $\exists\phi\in\Real
\suchthat
q - \phi p = 0$ are shaded in grey. Note that the only excluded points are
$p=0,\ q\in(-\infty,0)
\cup (0,\infty)$. $\ (q,p)=(0,0)$ is {\em not} excluded by the constraint.}
\label{toy_phasespace}
\end{figure}

The solution set can be thought of as an infinite dimensional generalization
of that shown in Fig. \ref{toy_phasespace}, which corresponds to the constraint
$\exists\phi\in\Real \suchthat
q - \phi p = 0$ on the classical state $(q,p)$ of a one degree of freedom
system.

Clearly (\ref{generalized_SD}), even though it contains the Lagrange multiplier
$\phi$, is more elegant than (\ref{content_SD}). Moreover, as shown in
Section \ref{Plebanski_action}, $\phi$ is the left-handed Weyl curvature (in
$SO(3)$ tensor
notation), which in the null initial value formulation of Lorentzian GR of
\cite{Penrose_Rindler1} contains the local degrees of freedom of the
gravitational field.
It therefore seems best to keep $\phi$ in the canonical theory.
At the end of this section a slightly different canonical formulation
will be given, in which $\phi$ is treated as a configuration variable. In
that formulation the constrained phase space takes on a more conventional,
manifold like, form.

Now we understand the constraint (\ref{generalized_SD}) a little better.
What about (\ref{Sigma_constraint})? And what is the significance of
$\tilE^a_i$ and $\Sgauge$?

To give some idea of what $\tilE^a_i$ and $\Sgauge$ represent I will evaluate
them in terms of tetrads on a slice, $\Sg_t$, of a non-degenerate solution
to the field equation (\ref{fieldeq1}), $\Sg_i\wedge\Sg_j\propto \dg_{ij}\eg$,
(which is equivalent to
(\ref{Sigma_constraint})). It was shown in Appendix \ref{sig_tet} that,
when (\ref{fieldeq1}) holds and $\Sg_i\wedge\Sg^i \neq 0$, $\Sg_i$ defines a
non-degenerate, orthonormal co-tetrad $e_\m^I$, unique up to $SO(3)_R$
transformations, so that
\be
\Sg_{i\,\m\n} = e^0_{[\m}e^i_{\n]} + \frac{1}{2}\eg_{ijk}e^j_\m e^k_\n.
\ee
Now the generators of $SO(3)_R$ are the anti-self-dual parts of the generators
of $SO(4)$, so $SO(3)_R$ transformations consist of an $SO(4)$ boost by an
arbitrary rapidity $\theta^i$, accompanied by a spatial rotation by an angle
$|\theta|$ about $\theta^i$. Hence $e^0_\m$ can be brought to any unit vector,
provided the rest of the tetrad is rotated appropriately. We will
take $e^0_\m = n_\m$, the future pointing unit normal to $\Sg_t$. Then
$e^0_a = 0$. Denoting the spatial co-triad $e^i_a$, in the adapted gauge,
by $E^i_a$ we find
\bearr
\Sg_{i\,ab} & = & \frac{1}{2}\eg_{ijk} E^j_a E^k_b	\\
	& = & \frac{1}{2}\eg_{abc} E^c_i\ det[E^i_a],
\eearr
where $E^a_i$ is the inverse of $E^i_a$. Applying the definition (\ref{defE})
one finds $\tilE^a_i = E^a_i\ det[E^i_a]$, showing that $\tilE^a_i$ is the
densitized spatial triad.\footnote{%
{\em Note added in proof:} $det[\tilE^a_i] = det[E^i_a]^2$, so real $\tilE$
with
$det\tilE <0$ correspond to pure imaginary $E^i_a$, and thus a negative
definite
spatial metric.}

Using this same gauge we can calculate $\Sgauge$ in terms of $\tilE^a_i$ and
the
lapse, $N$, and shift, $N^a$, defined by the time flow vector $v$ via
\be
v^\m = Nn^\m + N^\m, \ \ \ N^\m n_\m = 0.
\ee
Since $e^0_0 = n_\m v^\m = N$ and $e^i_0 = e^i_\m v^\m = e^i_\m N^\m = E^i_a
N^a,$
\bearr
\Sgauge & = & \frac{1}{2}N\,E^i_a + \frac{1}{2}\eg_{ijk}\,E^j_c E^k_a N^c \\
	& = & \frac{1}{4}\subtil{N}\eg_i{}^{jk}\eg_{abc}\,\tilE^b_j\tilE^c_k
+ \frac{1}{2}\eg_{abc}\tilE^b_i N^c.	\label{lapse_shift_form}
\eearr
Here $\subtil{N} = N/det[E^i_a].$

(\ref{lapse_shift_form}) can also be derived within the canonical theory from
(\ref{Sigma_constraint}), and the assumption that $\tilE^a_i$ is non-singular.
(\ref{Sigma_constraint}), $\Sg_{(i\,0a} \tilE^a_{j)} \propto \dg_{ij}$, implies
\be \label{Sig_soln}
\Sg_{i\,0a}\tilE^a_j = \tilde{c}_1 \dg_{ij} +  \tilde{c}_2^k\eg_{ikj},
\ee
where $\tilde{c}_1$ and $\tilde{c}_2^k$ are arbitrary densities.
If $rank\tilE = 3$ then (\ref{lapse_shift_form}) follows immediately from
(\ref{Sig_soln}) by contracting it
with the inverse of $\tilE$, $\subtil{E}_a^{-1\,i} = \frac{1}{2}
\eg^{ijk}\eg_{abc}\tilE^b_j\tilE^c_k/
det\tilE$, and setting $\subtil{N} = 2\tilde{c}_1/det\tilE$, and $N^c =
2\tilde{c}^k_2\tilE^c_k/
det\tilE$.

Solutions to (\ref{Sigma_constraint}) are still of the form
(\ref{lapse_shift_form}) where $rank\tilE = 2$
(see Appendix \ref{rank2app}).
However, where $rank\tilE \leq 1$ (\ref{lapse_shift_form}) is not the complete
solution. Rather, the general solution is
\be
\Sgauge = \frac{1}{2}\eg_{abc}\tilE^b_j N^{j\,c}_i,
\ee
which has two more degrees of freedom.
Finally, when $\tilE^a_i = 0$ $\Sgauge$ is completely unconstrained by
(\ref{Sigma_constraint}).

In \cite{CDJM} Capovilla, Dell, Jacobson and Mason derive Ashtekar's theory
from the Plebanski action by solving (\ref{Sigma_constraint}) (assuming
$rank\tilE
=3$) for $\Sgauge$, obtaining the lapse-shift form (\ref{lapse_shift_form}),
then substituting this form into the $3 + 1$ action (\ref{split_pleb2}).
Extremization of the action with respect to $N^c$ and $\subtil{N}$ then
yielded Ashtekar's constraints (\ref{vector_1}) and (\ref{scalar_1})
respectively. As the reader may easily verify, the constraints
(\ref{content_SD}) can be derived in the same way if the form of $\Sgauge$
appropriate to the rank of $\tilE$ is inserted into the action
(\ref{split_pleb2}).

The derivation of \cite{CDJM} leads to Ashtekar's Hamiltonian (\ref{H_Ash1}),
\be
H_{Ash} = - G_{A_0} - \int_\Sg s_{i\,a}(\subtil{N},\vec{N},\tilE)\tilB^{i\,a}\
d^3x,
\ee
where $s_{i\,a}$ is the lapse-shift form of $\Sgauge$, instead of the integral
of (\ref{Hdensity}),
\be
H_\Sg = - G_{A_0} - \int_\Sg \Sgauge [\tilB^{i\,a} - \phi^{ij}\tilE^a_j]\ d^3x.
\ee
In fact, when $\Sgauge = s_{i\,a}$, the two are equivalent.
Clearly $\{\tilE^a_i,H_\Sg\} = \frac{\dg H_\Sg}{\dg{A_a^i}}|_{\Sgauge =
s_{i\,a}}
= \{\tilE^a_i,H_{Ash}\}$. Less obviously
\be
\{A^i_a, H_{Ash}+ G_{A_0}\} = - \frac{\dg s_{j\,b}}{\dg\tilE^a_i}\tilB^{j\,b}
= - \frac{\dg s_{j\,b}}{\dg\tilE^a_i}\phi^{jk}\tilE^b_k = 0 + s_{j\,a}\phi^{ji}
= \{A^i_a, H_\Sg + G_{A_0}\}
\ee
by (\ref{generalized_SD}) and (\ref{Sigma_constraint}).

We now turn to proving the completeness of the constraints (\ref{Gauss_law})
and (\ref{generalized_SD}). To establish completeness we must show that the
constraints (\ref{Gauss_law}) and (\ref{generalized_SD}) are preserved
in the evolution, generated by the Hamiltonian, of any initial $(A,\tilE)$
satisfying (\ref{Gauss_law}) and (\ref{generalized_SD}).

Note that, since (\ref{generalized_SD}) requires only that there exists
$\phi^{ij}$, symmetric and traceless, such that $\tilB^{i\,a} - \phi^{ij}
\tilE^a_j = 0$, (\ref{generalized_SD}) is preserved by the evolution of the
state $(A, \tilE)$ provided there is a corresponding evolution of $\phi$
such that $\tilB^{i\,a} - \phi^{ij}\tilE^a_j$ remains zero. In other
words (\ref{generalized_SD}) is preserved if
\be 	\label{phi_evolution1}
0 = \frac{d}{dt}[\tilB^{i\,a} - \phi^{ij}\tilE^a_j] = \{\tilB^{i\,a}
- \phi^{ij}\tilE^a_j\, , H\} - \frac{d\phi^{ij}}{dt} \tilE^a_j
\ee
can be solved by some $\frac{d\phi^{ij}}{dt}$.

The Hamiltonian
is a sum of two parts proportional to the `constraint functions'
appearing in (\ref{Gauss_law}) and (\ref{generalized_SD}):
$H = H_1 + H_2$, with
$H_1 = -G_{A_0} = - \int_\Sg A^i_0 D_a \tilE^a_i d^3x$ and $H_2 = \int_\Sg
\Sg_{i\,0a}
[\tilB^{i\,a} - \phi^{ij}\tilE^a_j] d^3x$.

The Gauss law constraint (\ref{Gauss_law}), and thus $H_1$, generates $SO(3)$
gauge transformations. For infinitesimal $\Lam^i$
\bearr
\dg_\Lam A^i_a (x) & = & \{ A_a^i(x), \int_\Sg \Lam^j D_b \tilE^b_j d^3y\}
= - D_a \Lam^i (x) 	\label{dA_gauge}\\
\dg_\Lam \tilE_i^a (x) & = & \{ \tilE_i^a(x), \int_\Sg \Lam^j D_b \tilE^b_j
d^3y \} \\
& = & \{\tilE_i^a (x), -\int_\Sg \Lam^j \eg_{jk}{}^l  A_b^k\tilE^b_l d^3y\} \\
& = & -\Lam^j \eg_{ij}{}^k \tilE^a_k (x) \label{dE_gauge}
\eearr
is an infinitesimal $SO(3)$ gauge transformation.

The gauss law constraint (\ref{Gauss_law}) transforms homogeneously under
$SO(3)$ gauge transformations of $A$ and $\tilE$, so it is preserved by the
evolution generated by $H_1$.
That it is also preserved by $H_2$ can be seen as follows
\bearr
\{G_\Lam, H_2 \} & = &
\{ G_\Lam, -\int_\Sg \Sg_{j\,0b} [\tilB^{j\,b} - \phi^{jk}\tilE^b_k] d^3y\} \\
 	& = &  \int_\Sg \Sgauge [\dg_\Lam \tilB^{i\,a} - \phi^{ij}\dg_\Lam \tilE^a_j]
d^3x  \\
	& = & - \int_\Sg \dg_\Lam\Sgauge [\genSD] d^3x  \nonumber\\
	&   & \int_\Sg \dg_\Lam \phi^{ij}\Sg_{(i\,0a}\tilE^a_{j)} d^3x\\
\label{delH2gauss}
	& \approx & 0,
\eearr
where $\dg_\Lam$ now denotes the extension to all fields of the $SO(3)$ gauge
transformation generated by $G_\Lam$, and
$\approx 0$ means that the quantity vanishes on states solving the constraints.
The first
term in (\ref{delH2gauss}) vanishes when (\ref{generalized_SD}) holds, while
the second vanishes by virtue of the restriction (\ref{Sigma_constraint}) on
$\Sgauge$.

To check whether (\ref{generalized_SD}) is preserved we first compute
$\{\genSD, H\}$.
\be
\{\genSD, H_1\} \approx - \dg_{A_0}\phi^{ij}\tilE^a_j = - 2A_0^k\eg^{(i}{}_{kl}
\phi^{j)l} \tilE^a_j.
\ee
Now, for arbitrary $w^1_{i\, a}$ and
$w^2_{i\, a}$
\bearr
\lefteqn{\{ \int_\Sg w^1_{i\,a} [ \tilB^{i\,a} - \phi^{ij}\tilE^a_j]\ d^3x_1,
	\int_\Sg w^2_{k\,b} [\tilB^{k\,b} - \phi^{kl}\tilE^b_l]\ d^3x_2 \}}
\hspace{2cm}	\\
	& = &  \{ \int_\Sg w^1_{i\,a} \tilB^{i\,a}\ d^3x_1,
		  -	\int_\Sg w^2_{k\,b} \phi^{kl}\tilE^b_l\ d^3x_2 \}\ -\ 1 \exchange 2 \\
 	& = & - 2\int_\Sg [D_a w^1_{i\,b}] w^2_{j\,c}\eg^{abc} \phi^{ij} d^3x\ -\
1\exchange 2 \\
	& = & 2 \int_\Sg w^1_{i\,a}w^2_{j\,b}D_c \phi^{ij}\eg^{abc}\  d^3x,
\label{genSD_bracket}
\eearr
so
\be
\{\genSD, H_2\} = - 2\Sg_{j\,0b} D_c\phi^{ij} \eg^{abc}
\ee
(Note that the extremization of the action (\ref{split_pleb2}) with respect
to $\Sgauge$ requires the $\phi$ appearing in $H$ to be a $\phi$ which renders
$\genSD$ zero).

(\ref{phi_evolution1}) thus requires that\footnote{
This equation can also be derived directly in from the spacetime field
equations.
{}From $F^i - \phi^{ij}\Sg_j = 0$, $D\wedge\Sg_i = 0$ and the Bianchi identity
it
follows that
\be		\label{phi_fieldeq}
0 = D\wedge F^i - D\phi^{ij}\wedge\Sg_j - \phi^{ij}D\wedge\Sg_j = - D\phi^{ij}
\wedge\Sg_j.
\ee
Taking the $[0bc]$ component of this equation and contracting with $\eg^{abc}$
yields
\bearr
0  & = & 3 D_{[0}\phi^{ij}\Sg_{j\,bc]}\ \eg^{abc} \\
& & [D_0\phi^{ij} \Sg_{j\,bc} + D_c\phi^{ij} \Sg_{j\,0b} + D_b\phi^{ij}
\Sg_{j\,c0}]\eg^{abc} \\
& & D_0\phi^{ij} \tilE^a_j - 2 D_b\phi^{ij} \Sg_{j\,0c} \eg^{abc},
\eearr
i.e (\ref{phi_evolution2}).

The only other non-trivial component of (\ref{phi_fieldeq}), namely the
$[abc]$ component, requires $D_a\phi^{ij} \tilE^a_j = 0$, which holds
identically when (\ref{Gauss_law}) and (\ref{generalized_SD}) hold.}
\bearr
0 & = & \dot{\phi}^{ij}\tilE^a_j + 2A_0^k \eg^{(i}{}_{kl}\phi^{j)l}\tilE^a_j
	-2 D_b\phi^{ij} \Sg_{j\,0c}\eg^{abc}\\
  & = & D_0 \phi^{ij} \tilE^a_j - 2D_b\phi^{ij}\Sg_{j\,0c}\eg^{abc}.
\label{phi_evolution2}
\eearr

When $\Sgauge$ is of the lapse-shift form (\ref{lapse_shift_form})
(\ref{phi_evolution2}) can always be solved by some $D_0\phi^{ij}$. When
$\Sgauge$ is of this form
\be
\Sg_{j\,0c}\eg^{abc} = \frac{1}{2}\subtil{N}\eg_j{}^{kl}\tilE^a_k\tilE^b_l
+ \tilE^{[a}_j N^{b]}.
\ee
Hence (\ref{phi_evolution2}) becomes
\be
0 = [D_0\phi^{ij} - N^b D_b\phi^{ij} - \subtil{N} \eg_k{}^{jl}\tilE^b_l
D_b\phi^{ik} ]\tilE^a_j + D_b\phi^{ij}\tilE^b_j N^a.
\ee
The last term vanishes when (\ref{Gauss_law}) and (\ref{generalized_SD}) hold,
while the term in
brackets vanishes for suitable $D_0\phi^{ij}$, so the equation can be solved.

When $rank\tilE \geq 2$ (\ref{Sigma_constraint}) requires $\Sgauge$ to be of
the
lapse-shift form. Hence (\ref{phi_evolution2}) doesn't imply any new
restrictions on the Lagrange multipliers at a given time. In general the
solvability of (\ref{phi_evolution2}) requires that
\be
\exists \theta^{ij} \mbox{trace free, symmetric} \suchthat
2D_b\phi^{ij} \Sg_{j\,0c} \eg^{abc} = \theta^{ij}\tilE^a_j.  \label{exSgconst0}
\ee
(If this is true $D_0\phi^{ij} = \theta^{ij}$ solves (\ref{phi_evolution2})).
(\ref{exSgconst0}) is of the same form as (\ref{generalized_SD}). Its content
can be extracted by eliminating $\theta$. One finds
\be
0 = \tilE^{[d}_k\eg^{a]bc} D_b\phi^{ij} \Sg_{j0c}\ \ \mbox{if $rank\tilE = 1$,}
\label{Sg_const2}
\ee
and,
\be
0 = \eg^{abc} D_b\phi^{ij} \Sg_{j0c}\ \ \mbox{if $\tilE = 0$}.
\label{Sg_const3}
\ee
(The contraction on $k$ and $i$ of the right side of (\ref{Sg_const2}) vanishes
by (\ref{Gauss_law}), (\ref{generalized_SD}), and (\ref{Sigma_constraint}), but
the trace free
part produces new restrictions on $\Sgauge$).

For any $A$, $\tilE$, and $\phi$ the restrictions (\ref{Sg_const2}) and
(\ref{Sg_const3}) are solved by $\Sgauge$ of the lapse shift form, as well as
many others. Thus the preservation in time of (\ref{generalized_SD}) does not
require further, secondary constraints on $(A, \tilE)$, nor, in fact, on
$\phi$.

\section{Canonical formulation treating the Weyl curvature as a configuration
variable}
\label{three_plus_one2}

An illuminating alternative canonical formulation of the Plebanski theory
elevates $\phi^{ij}$ to the status of a configuration variable. This is
actually
a very natural thing to do. As pointed out in Section \ref{Plebanski_action},
$\phi^{ij}$ is really just the left-handed Weyl curvature (in $SO(3)$ tensor
language). In the null initial value formulation of Lorentzian GR of
\cite{Penrose_Rindler1}
a certain (complex) component of $\phi^{ij}$ constitutes the local
degrees of freedom of the gravitational field on the null initial surface.

$\phi^{ij}$ will thus be given a momentum $\tilpi_{ij}$ which will be
constrained to be zero. To keep the gauge invariance of the theory manifest,
the momentum $\tilpi$ will be `created' by adding a gauge invariant term to
the Plebanski action. The new action is
\be
I' = \int\frac{1}{2}\Sg_i\wedge F^i - \frac{1}{4}\phi^{ij}\Sg_i\wedge\Sg_j
+ \Pi_{ij}\wedge D\phi^{ij} + \Pi_{ij}\wedge K^{ij}.
\ee
The 3-form $\Pi_{ij}$ is symmetric and traceless in $ij$, and the 1-form
$K^{ij}$ is a Lagrange
multiplier which enforces $\Pi_{ij} = 0$. Note that the content of the
theory is completely unchanged, only the formalism describing it is being
modified.

A $3 + 1$ decomposition, and the definitions  $\tilpi_{ij} =
-\eg^{abc}\Pi_{ij\,abc}$ and $\kappa^{ij} = K_0^{ij}$, yields
\bearr
I' & = & \int\int_{\Sg_t} \tilE^a_i\di_0 A^i_a + \tilpi_{ij} D_0 \phi^{ij}
	+ A_0^i D_a\tilE^a_i + \Sgauge[\genSD] \\
	& & \ \ \ + \tilpi_{ij}\kappa^{ij} + 3\eg^{abc} \Pi_{ij\,0ab}
[D_c\phi^{ij} + K_c^{ij}]\ d^3x\ dt
\eearr
Extremization with respect to $K_c^{ij}$ requires $\Pi_{ij\,0ab} = 0$.
We may substitute this equation into the action and simply drop the last term.
Then we obtain a phase space action
which shows that the fundamental Poisson brackets are
\bearr
\{A^i_a(x), \tilE^b_j(y)\} & = & \dg^i_j\dg^a_b \dg^3(x,y)	\label{A_E_bracket}
\\
\{\phi^{ij}(x), \tilpi_{kl}(y)\} & = & [\dg^{(i}_k \dg^{j)}_l - \frac{1}{3}
\dg^{ij}\dg_{kl}] \dg^3(x,y), 	\label{phi_pi_bracket}
\eearr
the primary constraints are
\bearr
\caltilG'_i & = & D_a\tilE^a_i + 2\eg^j{}_{il}\phi^{lk}\tilpi_{jk} = 0
\label{Gauss_law2} \\
\caltilC^{i\,a} & = & \genSD = 0	\label{generalized_SD2} \\
\tilpi_{ij} & = & 0			\label{pi_0}
\eearr
and the Hamiltonian is
\be
H = - \int_\Sg A^i_0\ \caltilG'_i + \Sgauge\ \caltilC^{i\,a} + \kappa^{ij}\
\tilpi_{ij}\ d^3x.
\ee
$G_\Lam = \int_\Sg \Lam^i\caltilG'_i\ d^3x$ not only generates an $SO(3)$
gauge transformation of $A$ and $\tilE$, as shown in (\ref{dA_gauge}) and
(\ref{dE_gauge}), but also generates the corresponding gauge transformations
of $\phi$ and $\tilpi$:
\bearr
\{\phi^{ij}, G_\Lam \} & = & -2\Lam^k \eg^{(i}{}_{kl}\phi^{j)l} =
\dg_\Lam\phi^{ij}					\\
\{\tilpi_{ij}, G_\Lam \} & = & -2\Lam^k \eg_{(ik}{}^l\tilpi_{j)l} =
\dg_\Lam\tilpi_{ij}.
\eearr
The algebra of the integrated constraints $G_\Lam$, $C_w = \int_\Sg
w_{i\,a}\caltilC^{i\,a}\ d^3x$ and $P_u = \int_\Sg
u^{ij}\tilpi_{ij}\ d^3x$ now follows immediately from (\ref{phi_pi_bracket}),
(\ref{genSD_bracket})  and the fact that $G_\Lam$ generates the $SO(3)$ gauge
transformations $\dg_\Lam$:
\bearr
\{G_{\Lam_1}, G_{\Lam_2}\} & = & G_{[\Lam_1,\Lam_2]}	\label{GG_bracket} \\
\{G_\Lam, C_w\} & = & C_{\dg_\Lam\,w}			\label{GC_bracket} \\
\{G_\Lam, P_u\} & = & P_{\dg_\Lam\,u}			\label{GP_bracket} \\
\{C_{w^1}, C_{w^2}\} & = & 2 \int_\Sg w^1_{i\,a}w^2_{j\,b}\: D_c\phi^{ij}
\eg^{abc}\ d^3x						\label{CC_bracket} \\
\{C_w, P_u\} & = & - \int_\Sg w_{(i\,a}\tilE^a_{j)}\ u^{ij}\ d^3x
\label{CP_bracket} \\
\{P_{u_1}, P_{u_2}\} & = & 0				\label{PP_bracket}
\eearr
where $[\Lam_1,\Lam_2]^i = \eg^i{}_{jk}\Lam_1^j\Lam_2^k$ and $u^{ij}$ is
(without loss of generality) taken to be trace free.
(\ref{CC_bracket}) and (\ref{CP_bracket}) show that some of these constraints
are second class.

Nevertheless, when the restrictions on $\Sgauge$ found in our previous
$(A,\tilE)$ formulation of the canonical theory hold, the constraints are
preserved by evolution, so they are complete:
\bearr
\frac{d}{dt} G_\Lam & = & \{G_\Lam, H\} + G_{\frac{d\Lam}{dt}} \\
		& = & H[\dg_\Lam A^i_0, \dg_\Lam \Sg_{i\,0a}, \dg_\Lam \kappa^{ij}] +
G_{\frac{d\Lam}{dt}}\\
		& \approx & 0	\\
\frac{d}{dt} \caltilC^{i\,a} & \approx & \kappa^{ij} \tilE^a_j -
				2D_c\phi^{ij} \Sg_{j\,0b} \eg^{abc} \\
\frac{d}{dt} \tilpi_{ij} & \approx & - \Sg_{(i\,0a}\tilE^a_{j)} - \frac{1}{3}
\dg_{ij}\mbox{\em trace}.
\eearr
The constraints are preserved provided
\bearr
\Sg_{(i\,0a}\tilE^a_{j)} \propto \dg_{ij}	&&	\label{Sig_const1.2} \\
\kappa^{ij} \tilE^a_j - 2D_c\phi^{ij} \Sg_{j\,0b} \eg^{abc} = 0.	&&
\label{Sig_const2.2}
\eearr
(\ref{Sig_const1.2}) is of course just the familiar restriction
(\ref{Sigma_constraint}). Noting that the evolution equation of $\phi$ can
be written as $D_0\phi^{ij} = - \kappa^{ij}$ we see that (\ref{Sig_const2.2})
is just (\ref{phi_evolution2}):
\be
D_0\phi^{ij} \tilE^a_j + D_c\phi^{ij}\Sg_{j\,0b} \eg^{abc} = 0.
\ee
(\ref{Sig_const2.2}) can thus be solved for $\kappa^{ij}$ if and only if
$\Sg_{i\,0a}$
satisfies the conditions (\ref{Sig_const1.2}), (\ref{Sg_const2}) and
(\ref{Sg_const3}). If these conditions,
which place no constraints on the classical state $(A,\tilE,\phi,\tilpi)$,
hold, then
evolution preserves the primary constraints. Hence there are no secondary
constraints.\footnote{%
{\em Note added in proof}: In this theory the first class constraint subalgebra
consists
precisely of all Hamiltonians that preserve the constraints. Taking
$\Sg_{i\,0a}$ to
be of the lapse-shift form (\ref{lapse_shift_form}), which satisfies all
restrictions
on $\Sg_{i\,0a}$, and solving (\ref{Sig_const2.2}) for $\kappa$ we obtain first
class constraints
\bearr
\caltilS' & = & \frac{1}{4}\eg_i{}^{jk}\eg_{abc}\tilB^{i\,a}\tilE^b_j\tilE^c_k
				- \tilpi_{ij} \eg^{ik}{}_l \tilE^a_k D_a\phi^{lj}	\\
\caltilV'_a & = & \frac{1}{2}\eg_{abc}\tilB^{i\,b}\tilE^c_i - 2
\tilpi_{ij}D_a\phi^{ij}
\eearr
which are the extensions to the present phase space of the vector and scalar
constraints
of Ashtekar's theory. When $rank\tilE \geq 2$ the lapse shift form is the only
allowed
form of $\Sg_{i\,0a}$. $\caltilG'_i$, $\caltilV'_a$, and $\caltilS'$ then span
the
whole first class subalgebra. Furthermore, in this case the remaining (second
class)
constraints can be written as
\bearr
0 & = & C^{ij} = \phi^{ij} - f^{ij}(A,\tilE)	\\
0 & = & \tilpi
\eearr
In other words, the second class constraints simply fix $\phi$ and $\tilpi$ in
terms of
$A$ and $\tilE$. The Dirac bracket can be found in this case as follows.
$C^{ij}$,
$\tilpi_{ij}$, and $A'^i_a = A^i_a - \{A^i_a, \int_\Sg \tilpi_{kl} f^{kl} d^3x
\}$,
$\tilE'^a_i = \tilE^a_i - \{\tilE^a_i, \int_\Sg \tilpi_{kl} f^{kl} d^3x \}$ are
good
coordinates in a neighborhood of the constraint surface $C = \tilpi = 0$, while
at the
surface they are canonical coordinates, with $(C,\tilpi)$ being one canonically
conjugate
pair and $(A',\tilE')$ being the other. The Dirac bracket at the constraint
surface differs
from the Poisson bracket only in that $\{C,\tilpi\}_D = 0$, so that the only
non-zero
Dirac bracket of the coordinates is between $A'$ and $\tilE'$. Using the Dirac
bracket the
theory may be formulated completely on the constraint surface, with $C$ and
$\tilpi$ set
to zero once and for all. On this surface $A' = A$ and $\tilE' = \tilE$ form
canonical
coordinates and the theory is, in fact, identical to Ashtekar's.}

\begin{figure}
\centerline{\psfig{figure=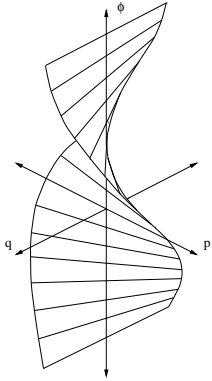,height=5cm}}
\caption{A part of the constraint surface $q - \phi p = 0$, $\pi = 0$.
The $\pi$ axis is not shown. The constraint manifold is a two dimensional
ruled surface, composed of the lines $q - \phi p = 0$ at every fixed
$\phi$. These lines rotate by $180^\circ$ from $\phi = -\infty$ to
$\phi = + \infty$. }
\label{toy_phasespace2}
\end{figure}

Let's consider the constraint surface of our second canonical formulation. An
analogous system with two degrees of freedom $q$ and $\phi$, with conjugate
momenta $p$ and $\pi$ respectively, is
\bearr
q - \phi p & = & 0		\label{one''}	\\
\pi & = & 0.			\label{two''}
\eearr
The phase space is four dimensional but (\ref{two''}) shows that the constraint
surface lies in the three dimensional subspace $\pi = 0$, so it can be
visualized. It is seen to be an infinite two dimensional plane which has been
twisted, like a ribbon, by a $180^\circ$ rotation of the  $\phi = + \infty$ end
relative
to the $\phi = -\infty$ end. (See Figure \ref{toy_phasespace2}). Note
that it is a manifold with no singularities.

Not surprisingly it is much harder to see what singularities the solution set
of the constraints (\ref{Gauss_law2}), (\ref{generalized_SD2}) and (\ref{pi_0})
has. However this much can be said. Since this solution set is the intersection
of the zeros of polynomials in the canonical variables it should not have any
cuts (i.e. excluded lower dimensional submanifolds) because it should, in some
sense, be a closed set.

The second class nature of the constraints poses a formidable obstacle to
canonical quantization. A short attempt did not yield any simple expression
for the Dirac bracket. The most promising approach, in the authors opinion,
is to take advantage of the simplicity of (\ref{generalized_SD2}) to
eliminate $A$ in favor of the other canonical variables. This seems
difficult at first, because $A$ cannot be expressed locally in terms of
$\tilB$ (and thus $\tilE$ and $\phi$ via (\ref{generalized_SD2})).

However, by integration, (\ref{generalized_SD}) can be turned into an
expression for the $SO(3)$ holonomies in terms of $\tilE$ and $\phi$ given in a
suitable
gauge. Thus (\ref{generalized_SD}) might be solvable if the gravitational
field is described in terms of holonomies (and some additional variables
to completely coordinatize phase space).

One might try to work either with the non-canonical classical loop variables
of Rovelli and Smolin \cite{RSloops90}, or with the canonical `Faraday
line' variables of Newman and Rovelli \cite{Newman_Rovelli92}, which describe
the ($SO(3)$ gauge
equivalence classes of) classical states as configurations of $\tilE$ flux
lines,
and fields canonically conjugate to those describing the flux lines.

The author hopes that ultimately solving (\ref{generalized_SD}) will
lead to a description of the gravitational field in terms of loops
and a dynamical $\phi$ field carrying the local degrees of freedom of
the field. Be that as it may, the problem of eliminating (\ref{generalized_SD})
will not be discussed further in this paper.



\section{Spacetime 2-sphere solution} \label{spt_sphere}

A `2-sphere solution' is a solution to the spacetime field equations
in which the basis, $\Sg_i$, of self-dual 2-forms, and the $SO(3)$ curvature,
$F^i$, both have support on an (unknotted) 2-sphere in spacetime, or, as is the
case in the present paper, on a thickened 2-sphere.

In Section \ref{covariance} it was argued that Ashtekar's canonical theory is
not fully {4-diffeomorphism} invariant because it does not have a 2-sphere
solution,
even though there is a 2-sphere spacetime field configuration which
solves the canonical theory on a suitable slicing of a neighborhood of any
point, and thus would be a solution if Ashtekar's theory were fully 4-diffeo
invariant. Here it will be shown that there is such a 2-sphere solution to the
spacetime field equations of Plebanski's theory (\ref{Plebanski_action}).
In Section \ref{canon_sphere} I will demonstrate that this spacetime field
configuration, viewed as a field history, also solves the canonical formulation
of Plebanski's theory that was worked out in Section \ref{three_plus_one}.
2-sphere solutions are especially interesting from the point of view of
canonical theory because they involve the birth and death of loops of
$\tilE$ and $\tilB$ flux. How, precisely, this comes about will be shown
in Section \ref{canon_sphere}.

We will begin with the ansatz
\be	\label{ansatz}
\Sg_i = e_i\sg \equiv e_i \int_0^1 d\lam_1\int_0^1 d\lam_2\ \Dg^*_{S_\lam} (x)
\ee
and then derive a corresponding $A^i$ such that the field equations
(\ref{fieldeq1}), (\ref{fieldeq2}) and (\ref{fieldeq3}) are solved.
$e_i$ is an $SO(3)$ vector field which will ultimately determine the internal
direction of the $\tilE^a_i$ in the canonical treatment. Without loss of
generality $e$ is taken to be non-zero. $\{S_\lam\}$ is a family of 2-spheres
parametrized by $\lam_A \in \Box \equiv [0,1]^2$. The $S_\lam$ do not intersect
each other, nor do they `bunch up' - the parameters $\lam_A$ are required to
be continuous functions on the part of spacetime occupied by the $S_\lam$.
$\Dg^*_{S_\lam\:\m\n} = \frac{1}{2}\eg_{\m\n\sg\rho} \Dg^{\sg\rho}_{S_\lam}$
is the spacetime dual\footnote{
The spacetime dual of an $n$-form $g$ will be defined as
\be
g^{*\,\m_{n+1}...\m_d} = \frac{1}{n!} \eg^{\n_1...\n_n\m_{n+1}...\m_d}
g_{\n_1...\n_n},   \label{form_dual}
\ee
and that of an $m$-vector, $h$, as
\be
h^*_{\m_1...\m_{d-m}} = \frac{1}{m!} \epsilon_{\m_1...\m_{d-m}\n_1...\n_m}
h^{\n_1...\n_m}.
\label{vector_dual}
\ee
With these definitions $** = 1$.
Note that $*$ has nothing to do with a metric. Both $\eg_{\m_1...\m_d}$
and $\eg^{\m_1...\m_d}$ are antisymmetric symbols with
$\eg_{12...d}=\eg^{12...d}=1$.}
of the characteristic distribution of $S_\lam$:
$\Dg_{S_\lam}^{\m\n}(x) = \int_{S_\lam} \dg^4(x - z)\:dz^{[\m}dz^{\n]}$.
$\Dg_S^{\m\n}$ is a generalization to 2-surfaces of the current of a worldline.
It can be thought of as a second degree delta function with support on $S$,
times the local tangent bivector of $S$. More on characteristic distributions
can be found in Appendix \ref{characteristic_dist}.

In (\ref{ansatz}) $\Sg_i$ is supported on a 2-sphere thickened in two
dimensions, i.e. on a 4-volume $\calS = \cup_{\lam \in \Box} S_\lam$. This has
the
advantage that the fields are regular enough that the theory of Sections
\ref{Plebanski_action} and \ref{three_plus_one} can be applied without
modification.\footnote{
It seems that one can actually get away with thickening the 2-sphere in only
one dimension. However, to accommodate $\Sg_i$ with support strictly on a
2-surface requires an extension of Plebanski's theory, because in this case,
according to (\ref{fieldeq1}), $F^i$ would also have support strictly on the
2-surface. Such an $F^i$ cannot be defined without framing the surface
because the holonomy of a loop around the 2-surface depends on its base point
even when the loop shrinks to a point. (A framing would define a base point
for all infinitesimal loops around the surface). Similarly, parallel transport
on the 2-surface, which we will see is essential for defining solutions,
requires a framing to define which paths `wind around the surface' and which
do not.
Perhaps this is the source of the problems encountered by Bostr\"om, Miller
and Smolin in their attempt \cite{Bos94} to construct an analogue of Regge
calculus using $\Sg_i$ supported on 2-surfaces.}

Now let's find the consequences of each of the field equations in turn within
the ansatz (\ref{ansatz}).

(\ref{fieldeq1}) requires $\Sg_i\wedge\Sg_j \propto \dg_{ij}\eg$. According to
(\ref{ansatz})
\be
\Sg_i\wedge\Sg_j = e_i e_j \sg\wedge\sg.
\ee
Note that the independent tangents to $S_\lam$, $t_1$ and $t_2$, satisfy
$t_A^{[\m} \Dg_{S_\lam}^{\n]\sg} = 0$, which implies $t_A^\m \Dg^*_{S_\lam\:
\m\n} = 0$. Since $\lam_A$ are continuous on $\calS$ a unique $S_\lam$
passes through each point of $\calS$. Hence $t_1^\m$ and $t_2^\m$ are
spacetime vector fields with the property $t_A^\m\sg_{\m\n} = 0$ on $\calS$,
which is the support of $\sg$.
$\sg\wedge\sg = 0$ follows immediately, and thus $\Sg_i\wedge\Sg_j = 0$,
implying that (\ref{fieldeq1}) holds identically in the ansatz (\ref{ansatz}).

(\ref{fieldeq2}) requires $D\wedge\Sg_i = 0$, or equivalently $D_\m
\Sg^{*\:\m\n}_i = 0$. According to (\ref{ansatz})
\be	\label{div_decomposition}
D_\m  \Sg^{*\:\m\n}_i = D_\m e_i \sg^{*\:\m\n} + e_i \di_\m \sg^{*\:\m\n}.
\ee
$** = 1$ and (\ref{boundary_id}) then show that
\be
\di_\m \sg^{*\:\m\n} = \int_\Box d^2\lam\ \di_\m\Dg^{\m\n}_{S_\lam}
= - \frac{1}{2} \int_\Box d^2\lam\ \Dg^{\n}_{\di S_\lam}.
\ee
Since the $S_\lam$ are 2-spheres $\di S_\lam = \emptyset$ which means that the
second term in (\ref{div_decomposition}) vanishes.
(\ref{fieldeq2}) thus requires
\be  	\label{f2_imp}
D_\m e_i \sg^{*\,\m\n} = 0.
\ee
But $\sg^{*\,\m\n} \propto t_1^{[\m} t_2^{\n]}$ (the tangent bivector of the
$S_\lam$ passing through the point), so (\ref{f2_imp}) implies that
$t_A^\m D_\m e_i = 0$, i.e. that $e$ is covariantly constant on the $S_\lam$!

(\ref{fieldeq3}) requires $F^i - \phi^{ij}\Sg_j = 0$. (\ref{ansatz}) then
implies that
\be 	\label{f3_imp}
F^i = \phi^{ij} e_j \sg = b^i\sg.
\ee
$F^i$ also has support on $\calS$, and $t_A^\m F^i_{\m\n} = 0$. The connection
is thus flat on each $S_\lam$. Since the $S_\lam$, being 2-spheres, have no
non-contractable curves, the condition that $e$ be covariantly constant on
$S_\lam$ can be solved because parallel transport is completely path
independent on $S_\lam$.\footnote{
If the $S_\lam$ had higher genus $\calS$ could thread through handles in
$S_\lam$. The curvature $b^i \sg$ would then induce a non-trivial holonomy
around non-contractable curves.}

Hence, to find a solution to all the field equations with $\Sg_i$ of the form
(\ref{ansatz}) one only needs to find a connection, $A^i$, having curvature
$F^i = b^i\sg$, with $b^i - \phi^{ij} e_j = 0$ and $e_i$ covariantly constant
on $S_\lam$.

The Bianchi identity,
\be
D\wedge F^i = 0,
\ee
requires (in analogy to (\ref{fieldeq2})) that $t_a^\m D_\m b^i = 0$. $b$ must,
like $e$, be covariantly constant on $S_\lam$. $b$ is, however, not further
constrained by the requirement $b^i - \phi^{ij} e_j = 0$.
For arbitrary $SO(3)$ vectors $b$ and $e$ ($e\neq 0$) this requirement is met
by
\be
\phi^{ij} = \frac{1}{e_1}\left[ \begin{array}{ccc}
	b^1 & b^2 & b^3 \\
	b^2 & \theta - b^1 & \varphi  \\
	b^3 & \varphi & -\theta
	\end{array} \right]
\ee
where the internal $1$ axis has been taken to lie along $e$. The degrees of
freedom $\theta$ and $\varphi$ can be set arbitrarily at every point without
affecting any other fields in the solution. They can, and will, be set to zero.
Then $\phi$ too is covariantly constant on the $S_\lam$.

Beyond the Bianchi identity the existence of an $A^i$ poses no further
restrictions on $F^i = b^i \sg$. $A^i$ is easily found in the gauge in which
$e$ has constant components on each $S_\lam$, i.e. $e_i = e_i (\lam)$,
and $b$ has constant components on all of $\calS$. Since the $S_\lam$ are
closed
there exist 3-manifolds $U_\lam$ such that $\di U_\lam = S_\lam$\footnote{%
We are assuming that the $S_\lam$ are not non-contractable 2-spheres. If we
assume that the spacetime has topology $S^3\times\Real$, as we did in Section
\ref{three_plus_one}, then there are no non-contractable 2-spheres.}
Moreover, the $U_\lam$ may be chosen so that they do not cut any $S_\lam$
transversely. That is to say, if $S_\lam$ touches $U_{\lam'}$ then it lies
entirely in $U_{\lam'}$. Now let
\be
A^i = b^i \ag \equiv 3! b^i \int_\Box d^2\lam\ \Dg^*_{U_\lam},
\ee
and set the components of $b$ constant on all spacetime.
Then
\be
F^i = d\wedge A^i + \eg^i{}_{jk} A^j\wedge A^k = b^i d\wedge\ag + \ag\wedge\ag
\:b^j b^k \eg^i{}_{jk} = b^i d\wedge\ag.
\ee
{}From (\ref{boundary_id}) of Appendix \ref{characteristic_dist}
\be
d\wedge\ag = 3! \int_\Box d^2\lam\ d\wedge \Dg^*_{U_\lam} = \int_\Box d^2\lam\
\Dg^*_{S_\lam} = \sg,
\ee
so indeed $F^i = b^i\sg$.

The $U_\lam$ have been chosen so that the tangents, $t_A^\m$, to the $S_\lam$
are also tangent to $U_\lam$. It follows that $t_A^\m A_\m^i = 0$, i.e.
the connection components along $S_\lam$ vanish, which shows that $e_i =
e_i(\lam)$ and $b^i$ are covariantly constant on the $S_\lam$. We have found
the solution corresponding to ansatz (\ref{ansatz})!

This solution can be stated, in the same gauge, with less emphasis on the
2-surfaces $S_\lam$:
\bearr
A^i_\m & = & b^i\ag_\m			\label{sphere_A} \\
\Sg_{i\,\m\n} & = & e_i\sg_{\m\n}	\label{sphere_Sg} \\
\sg & = & d\wedge\ag			\label{sg_da} \\
db_i & = & 0 				\label{b_constant} \\
\sg\wedge de_i & = & 0.			\label{e_const_on_S}
\eearr
The field $\ag$ satisfies the condition
\be
\ag\wedge\sg  =  0.			\label{a_sg}
\ee
The exterior derivative of (\ref{a_sg}) is
\be
\sg\wedge\sg  =  0			\label{sg_sg}
\ee
which shows that there exist linearly independent vector fields $t_1^\m$,
$t_2^\m$ such that $t_A^\m\sg_{\m\n} = 0$. (\ref{sg_da}) implies that
$(t_1,t_2)$ integrate to form surfaces. These are 2-spheres in the 2-sphere
solution. Beyond (\ref{sg_sg}) (\ref{a_sg}) implies the gauge condition
$t_A^\m\ag_\m = 0$. Note finally that (\ref{e_const_on_S}) is equivalent
to $t_A^\m \di_\m e_i = 0$, so $e_i$ is constant on the integral surfaces
of the $t_A$.

Given (\ref{sphere_A}) - (\ref{e_const_on_S}) it is easy to show that the
field equations hold.


\section{Canonical form of the 2-sphere solution} \label{canon_sphere}

In Section~\ref{spt_sphere} a `2-sphere' solution to Plebanski's
spacetime field equations was found. Here  we verify that the corresponding
histories of canonical fields solve the canonical formulation of Plebanski's
theory given in Section \ref{three_plus_one}. Then the evolution of canonical
fields, especially the birth process, in a simple slicing $\Sg_t$ is studied in
detail. For clarity only solutions with $e_i$ constant are treated. The
analysis
extends easily to $e$ depending on $\lambda$.

As a first step the 2-sphere solution of Section~\ref{spt_sphere} will be
restated as a history of canonical field configurations on $\Sg$, and the
constraints, restrictions on the Lagrange multipliers, and evolution equations
verified. Then the evolution prior to birth, during life, and especially
during birth will be examined in detail.

The definition (\ref{sphere_A}) - (\ref{e_const_on_S})  of the 2-sphere
solution, and the specialization $e_i = \mbox{\em constant}$ will be taken as
the starting point for the translation into
canonical language. Thus, on $\Sg$,
\bearr
A^i_a & = & b^i \ag_a    	\label{csphere_A}\\
\tilE^a_i & = & e_i \tilw^a	\label{csphere_E}
\eearr
where $b^i$ and $e_i$ are constant $SO(3)$ vectors and we have
defined $\tilw^a = \eg^{abc}\sg_{bc}$.

The Lagrange multipliers are given by
\bearr
A^i_0 & = & b^i\ag_0 		\label{csphere_A0}\\
\Sg_{i\,0a} & = & e_i \sg_{0a}	\label{csphere_Sg}\\
b^i - \phi^{ij} e_j & = & 0		\label{csphere_phi}
\eearr
with $\phi$ traceless, symmetric and constant. (Such a $\phi$ exists for all
choices of constant $e$ and $b$).

(\ref{sg_da}), (\ref{sg_sg}), and (\ref{a_sg}) imply the following restrictions
on $\ag_0$, $\ag_a$, $\sg_{0a}$, and $\tilw^a$ on $\Sg$:
\bearr
\tilw^a & = & 2 \eg^{abc} \di_b \ag_c   \label{w_da}\\
0 & = & \tilw^a \sg_{0a} 		\label{w_sg}\\
0 & = & \ag_a \tilw^a			\label{a_w}\\
0 & = & \ag_0 \tilw^a + 2 \eg^{abc}\ag_c \sg_{0b}.	\label{a0_w}
\eearr
Notice that (\ref{w_da}) implies that
\be
\di_a \tilw^a = 0. 			\label{div_w}
\ee
Hence, $\tilw^a$ defines closed `field lines'.
These field lines are just the intersections $S_\lam\cap \Sg_t$ where
$S_\lam$ cuts $\Sg_t$ transversely.

(\ref{sg_da}) also yields an evolution equation for $\ag_a$:
\be
\sg_{0a} = \dot{\ag}_a - \di_a \ag_0.	\label{a_evolution}
\ee
(\ref{w_da}) and the gradient of (\ref{a_evolution}) in turn give an evolution
equation for $\tilw^a$:
\be
\dot{\tilw}^a = 2\eg^{abc} \di_b \sg_{0c}.	\label{w_evolution}
\ee

The constraints, restrictions on Lagrange multipliers, and evolution equations
of the canonical theory can now be shown to hold using (\ref{csphere_A}) -
(\ref{w_evolution}). First the constraints.
\be
D_a \tilE^a_i = e_i\,\di_a\tilw^a + \eg_{ij}{}^k b^j e_k\,\ag_a \tilw^a = 0
\ee
by (\ref{a_evolution}) and (\ref{a_w}). Constraint (\ref{Gauss_law}) holds.
\bearr
\tilB^{i\,a} & = & 2\eg^{abc}\,[b^i \,\di_b \ag_c + \eg^i{}_{jk}b^j b^k\
\ag_b\ag_c] \\
	& = & b^i \tilw^a
\eearr
by (\ref{w_da}), so $\tilB^{i\,a} - \phi^{ij}\tilE^a_j = [b^i - \phi^{ij}e_j]
\tilw^a = 0$. Thus constraint (\ref{generalized_SD}) holds.

The Lagrange multiplier $\Sg_{i\,0a}$ satisfies
\be
\Sg_{(i\,0a}\tilE^a_{j)} = e_{(i}e_{j)}\ \sg_{0a}\tilw^a = 0
\ee
by (\ref{w_sg}). Thus it satisfies (\ref{Sigma_constraint}). $\phi^{ij}$ must
satisfy
\bearr
0 & = & D_0 \phi^{ij}\ \tilE^a_j + 2\eg^{abc} D_c\phi^{ij}\ \Sg_{j\,0b}\\
  & = & \dot{\phi}^{ij}\ e_j\tilw^a + 2\eg^{abc}\ \di_c\phi^{ij} e_j\
\sg_{0b}\\
&& + 2\eg^{(i}{}_{kl}\phi^{j)l} b^k e_j\ [\ag_0\tilw^a +
2\eg^{abc}\ag_c\sg_{ob}]\\
& = & 0.
\eearr
This holds because of (\ref{a0_w}) and because $\phi$ is constant.

The Lagrange multipliers obey all the restrictions they should. There remains
to check the evolution equations,
\bearr
\dot{A}^i_a & = & D_a A^i_0 + \phi^{ij} \Sg_{j\,0a} \label{A_evolution} \\
\dot{E}^a_i & = & -\eg_{ij}{}^k A_0^j \tilE^a_k + 2\eg^{abc} D_b\Sg_{i\,0c}.
\label{E_evolution}
\eearr
In the 2-surface solution the right side of (\ref{A_evolution}) is
\bearr
\lefteqn{b^i\ \di_a\ag_0 + \eg^i{}_{jk} b^j b^k\ \ag_a\ag_0 + \phi^{ij} e_j\
\sg_{0a}}  \hspace{2cm}\\
	& = & b^i\ [\di_a\ag_0 + \sg_{0a}]\\
	& = & b^i\dot{\ag}_a \\
	& = & \dot{A}^i_a
\eearr
by (\ref{a_evolution}), so (\ref{A_evolution}) holds.
The right side of (\ref{E_evolution}) is
\bearr
\lefteqn{2 e_i\ \eg^{abc}\ \di_b\sg_{0c} + 2 \eg_{ij}{}^k b^j e_k\
\eg^{abc}\ag_b\sg_{0c}
- \eg_{ij}{}^k b^j e_k\ \ag_0\tilw^a} 	\hspace{2cm}	\\
& = & 2 e_i\ \eg^{abc}\di_b\sg_{0c} - \eg_{ij}{}^k b^j e_k\ [\ag_0\tilw^a +
2\eg^{abc}\ag_c\sg_{0b}] \\
& = & e_i \dot{\tilw}^a\\
& = & \dot{\tilE}^a_i.
\eearr
The evolution equations hold.

We have shown, in a somewhat abstract way, that
the 2-sphere solution is indeed a solution to the canonical theory developed in
Section \ref{three_plus_one}. Now let's choose a particular slicing and try to
understand more intuitively what happens before the flux lines are born,
during birth, and how, once born, the flux lines evolve.

\begin{figure}
\centerline{\psfig{figure=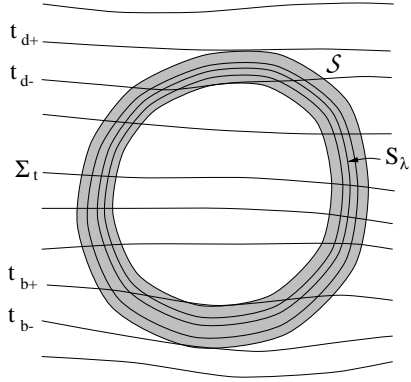,height=5cm}}
\caption{The support $\calS$ of $\Sg_i$ and $F^i$ in the
2-sphere solution is shown schematically, as are the 2-spheres $S_\lam$
and the slicing $\Sg_t$ used in the discussion of the phases of evolution
of the 2-sphere solution.}
\label{sphere_slicing}
\end{figure}

We will use a simple slicing, $\Sg_t$, in which $\calS_t = {\cal S}\cap \Sg_t$
is $\emptyset$
for $t<t_{b-}$, then becomes a simply connected ball until $t_{b+}$ when it
turns into a torus, which expands, recontracts and turns back into a ball
at $t_{d-}$, and disappears altogether at $t_{d+}$, so that $\calS_t=
\emptyset$ for $t > t_{d+}$. Figure~\ref{sphere_slicing} illustrates
the significance of $t_{b-}$, $t_{b+}$, $t_{d-}$, and $t_{d+}$. The time
interval from $t_{b-}$ to $t_{b+}$ will be called ``birth", $(t_{b+}, t_{d-})$
will be called ``life", and $(t_{d-},t_{d+})$ will be called ``death".

The slicing will also be required to be such that the 2-spheres
$S_\lam$ that fiber $\cal S$ do not ``go back and forth in time". In other
words, $t$ has only a minimum (during birth) and a maximum (during death), and
no other stationary points on $S_\lam$. (However, the maximum and minimum will,
in general, be allowed to occupy open subsets of $S_\lam$). Except at
stationary points of $t$ on $S_\lam$ the tangents $t_A^\m$ of $S_\lam$ are not
both spatial,
so the last condition implies that for $t_{min}<t<t_{max}$ on $S_\lam$
$t_A^0 \neq 0$ for $A = 1$ or $2$.

The pre-birth phase is quite featureless. Since $\calS_t = \emptyset$
for $t<t_{b-}$, $\tilw^a = 0$, so there is no $\tilE$ or $\tilB$ field.
The evolution, which consists purely of $SO(3)$ gauge transformations of the
pure gauge $A^i_a$ field, is generated by the Hamiltonian $H_o = -\int_\Sg
A^i_0 D_a \tilE^a_i$.

During the lifetime $\tilE$ and $\tilB$ also evolve quite straightforwardly.
$\tilw^a$ is a divergenceless vector density field living in the torus
$\calS_t$, defining `field lines' filling this torus. One of the tangent
vectors to $S_\lam$, say $t_2$, may be taken to be spatial, i.e. to lie along
the intersection $\cg_{\lam,t} = S_\lam\cap\Sg_t$.
But then $0 = t_2^\m\sg_{\m b} = t_2^a\sg_{ab} =
\frac{1}{2}\eg_{abc}t_2^a\tilw^c$,
implies that $\tilw^a$ lies along $t_1^a$. The `field lines' of $\tilw$ exactly
trace
the intersection curves $\cg_{\lam,t}$. In fact
\be
\tilw^a = \int_\Box d^2\lam\ \Dgthree^a_{\cg_{\lam,t}} (x)	\label{w_fieldlines}
\ee
where $\Dgthree^a_\cg = \int_\cg \dg^3(x-z)\ dz^a$ is the characteristic
distribution of the curve $\cg$, in three dimensions.

\noindent Proof:
\be
\tilw^a (x) = \eg^{abc}\sg_{bc} = 2 \int_\Box d^2\lam\ \Dg^{0a}_{S_\lam}(x,t)
\ee
$\Dg^{0a}_{S_\lam}(x,t) = \int_{S_\lam} \dg(t-z^0)\dg^3(x-z)\ \frac{\di z^{[0}}
{\di \sg^1}\frac{\di z^{a]}}{\di\sg^2}\ d^2\sg$, where $\sg^A$ are coordinates
on $S_\lam$. By choosing $\sg^1 = z^0$ (which can be done since the time,
$z^0$,
has no stationary points on $S_\lam$ during `life') we see that
$\Dg^{0a}_{S_\lam} = \frac{1}{2}\int_{\cg_{\lam,t}} \dg^3(x - z) dz^a
= \frac{1}{2}\Dgthree^a_{\cg_{\lam,t}}$, and (\ref{w_fieldlines}) follows.\QED

Keeping the choice of coordinate $\sg^1 = z^0$ on $S_\lam$ take as $t_1\ $
$\frac{\di}{\di\sg^1}$. Then $t_1^0 = 1$ and $\sg_{0a} = -t_1^b\sg_{ba}
= -\frac{1}{2}\eg_{abc}\tilw^b t_1^c$. This gives us a $\Sg_{i\,0a}$ of the
lapse-shift form (\ref{lapse_shift_form}):
\be
\Sg_{i\,0a} = e_i\sg_{0a} = \frac{1}{2}\eg_{abc}\tilE^b_i N^c,
\ee
where $N^c = -t_1^c$.\footnote{
In non-degenerate solutions the shift is $N^a = -n^a/n^0$, where $n^\m$ is the
unit normal to $\Sg_t$. In the degenerate solution we are considering $n$
is not well defined, but we see that, in a sense, $t_1$ is `normal' to
$\Sg_t$.}

Evolution is thus generated by
\be
H_l = - \int_\Sg A_0^i D_a \tilE^a_i + \frac{1}{2}\eg_{abc}\tilB^{i\,a}
\tilE^b_i N^c\ d^3x,
\ee
which, like $H_o$, is a special case of the Ashtekar Hamiltonian.\footnote{
Recall that when $\Sg_{i\,0a}$ is of the lapse-shift form
$s_{i\,a}(\subtil{N},\vec{N},\tilE)$ one may calculate evolution from either,
the Hamiltonian density
\be
\tilde{\cal H}_\Sg = -A_0^i D_a\tilE^a_i - \Sg_{i\,0a}[\tilB^{i\,a} - \phi^{ij}
\tilE^a_j ],
\ee
substituting $\Sg_{i\,0a} = s_{i\,a}(\subtil{N},\vec{N},\tilE)$ into the
evolution equation {\em after} the Poisson brackets have been evaluated, or
the lapse-shift form
\be
\tilde{\cal H}_N = -A_0^i D_a\tilE^a_i -
s_{i\,a}(\subtil{N},\vec{N},\tilE)\tilB^{i\,a},
\ee
which is just Ashtekar's Hamiltonian density.}

$\tilE$ and $\tilB$ evolve only by spatial diffeomorphisms, as can be seen
by evolving with $H_l$ or, more simply, from the evolution (\ref{w_evolution})
of $\tilw$.
\bearr
\dot{\tilw}^a & = & 2\ \di_b\sg_{0c}\ \eg^{abc} \\
	& = & 2\di_b [\tilw^{[a} N^{b]}]     \\
	& = & N^b\di_b\tilw^a + \tilw^a\di_b N^b - \di_b N^a \tilw^b \\
	& = & \pounds_{\vec{N}}\ \tilw^a,
\eearr
which then implies $\dot{\tilE}^a_i = \pounds_{\vec{N}}\:\tilE^a_i$ and
$\dot{\tilB}^{i\,a} = \pounds_{\vec{N}}\:\tilB^{i\,a}$, since $e_i$ and $b^i$
are
constant.

The most interesting aspect of the 2-sphere solution is the birth. During the
birth there are points in $\calS_t$ at which
an $S_\lam$ touches $\Sg_t$ tangentially, and thus both $t_1$ and $t_2$ are
spatial.

\begin{figure}
\centerline{\psfig{figure=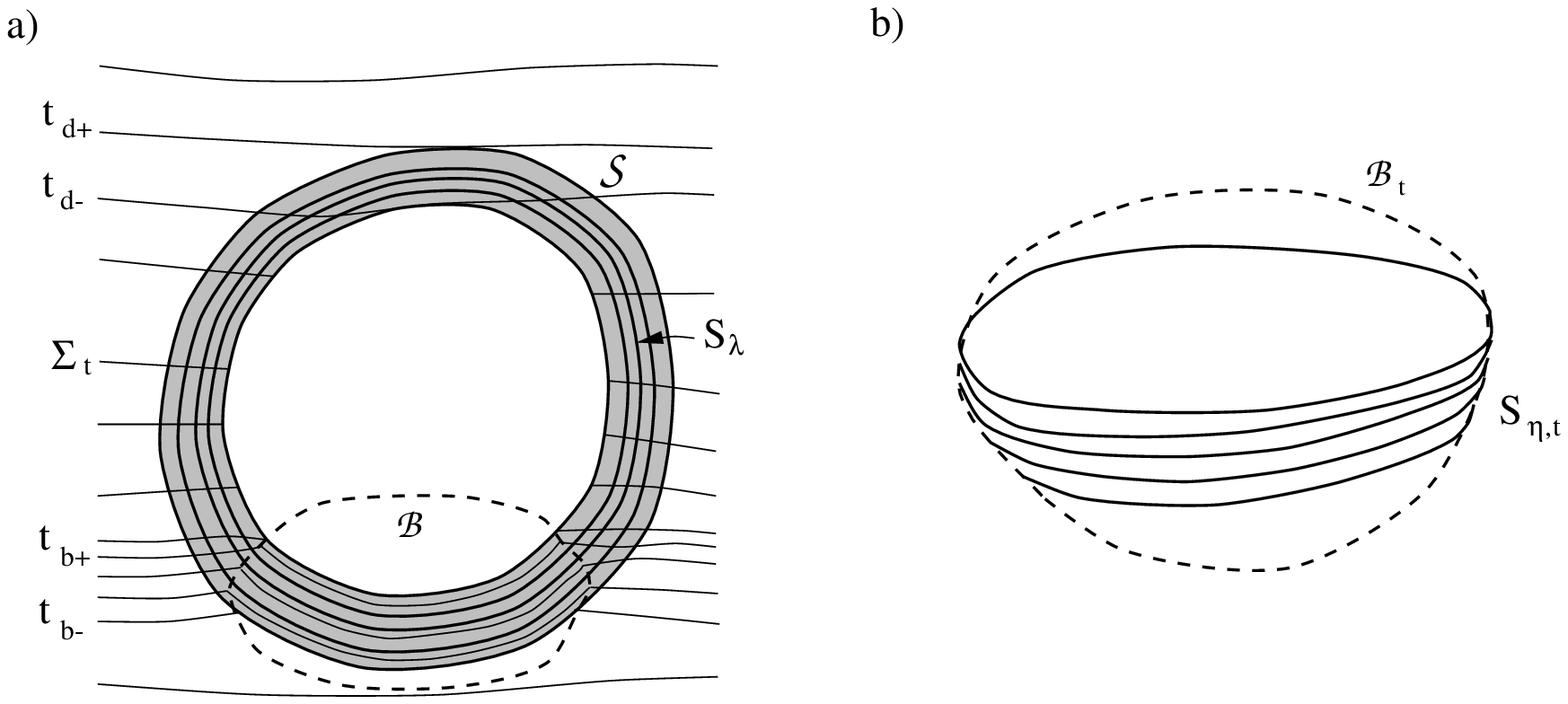,height=5cm}}
\caption[yyy]{Panel a) shows the special slicing in which the initial equal
time slices of the 2-spheres $S_\lam$ are finite patches $S_{\eta,t}$ of
2-surface,
and consequently flux loops are born with finite size.

Panel b) shows the 3-volume $\calB_t = \calB\cap\Sg_t$ and the patches
$S_{\eta,t}$
of the $S_\lam$ in $\calB_t$. }
\label{special_slicing}
\end{figure}

Generically such points form a line in $\Sg_t$, but, by chosing a suitable
slicing, $t_1$ and $t_2$ can be made spatial in the slices,
$\calB_t = {\cal B}\cap \Sg_t$,
of an open set $\cal B$ in spacetime. For conceptual simplicity let us assume
for the moment that such a slicing has been chosen. Then, in ${\cal B}_t$,
$\tilw^a = 0$, while $\sg_{0a} \neq 0$ but is proportional to $\eg_{abc}t_1^b
t_2^c$. In fact $\sg_{0a}$ is just the dual of the average over $\lam_A$ of the
characteristic distributions of the $S_\lam$, which are tangent to the $\Sg_t$
in $\cal B$. That is, in ${\cal B}_t$
\bearr
\sg_{0a}(x) = \bar{\sg}_{0a}(x) & = & \frac{1}{2}\eg_{abc} \int_\Box d^2\lam
\ \Dg_{S_\lam\cap{\cal B}}^{bc} (x,t) \\
& = & \frac{1}{2} \int_\Real d\eta\ f(\eta)\ \Dgthree^{bc}_{S_{\eta,t}}(x).
\eearr
Here $\eta = g(\lam)$ is chosen so that $\eta$ and $t_{min}$, the minimum
of $t$ on a surface, parametrize the surfaces $\{S_\lam\}$, $S_{\eta,t} =
S_{\lam(\eta,t_{min})}\cap \calB_t$,
and $f(\eta) = \int_\Box d^2\lam\ \dg(t-z^0(\lam))\dg(\eta - g(\lam))$
is essentially a Jacobian. Note that in $\calB$ $\ z^0$ depends only on $\lam$,
since the $S_\lam$ are tangent to the $\Sg_t$ there.
Figure~\ref{special_slicing} illustrates the slicing and the $S_{\eta,t}$.

In $\calB_t$ $\ \Sg_{i\,0a} = \bg_{i\,a} \equiv e_i \bar{\sg}_{0a}$, which
is not of the lapse-shift form (\ref{lapse_shift_form}), and hence contributes
a term to the Hamiltonian,
\be
\Dg H_b = -\int_\Sg \bg_{i\,a}\tilB^{i\,a},
\ee
which is not present in $H_{Ash}$. Note that $\Dg H_b$ is independent of
$\tilE$,
which is zero in $\calB_t$.\footnote{
As will be seen $H_b$ generates the birth of loops of $\tilE$ and $\tilB$
field. The $e_i$ appearing in $\bg_{i\,a}$ sets the internal
$SO(3)$ direction of the $\tilE$ field that is generated. It does {\em not}
indicate a dependence of $\bg$ on the existing $\tilE$ field, which is zero.}

Outside $\calB_t$ $\ \Sg_{i\,0a}$, and thus the Hamiltonian density, is of
the same form as in the `life' interval $(t_{b+},t_{d-})$. The total
Hamiltonian
during the birth is thus
\be
H_b = \Dg H_b + H_l.
\ee

Now let's consider evolution during the birth. $\Dg H_b$ contributes to the
evolution of $\tilE$
\bearr
\Dg\dot{\tilE}^a_i & = & 2\eg^{abc}D_b\bg_{i\,c} \\
	& = & 2\eg^{abc} e_i\ \di_b\bar{\sg}_{0c} + 2\eg^{abc} D_b e_i\
\bar{\sg}_{0c}   \label{DgEdot_expression1} \\
	& = & 2 \int d\eta\ f(\eta)\ [e_i\di_b\ \Dgthree^{ab}_{S_{\eta,t}} +
D_b e_i\ \Dgthree^{ab}_{S_{\eta,t}}]	\label{DgEdot_expression2} \\
	& = &  \int d\eta\ f(\eta)\ e_i\ \Dgthree^a_{\di S_{\eta,t}}.
\eearr
$\Dg H_b$ thus generates the birth of $\tilE$ field lines $e_i\ \Dgthree^a_{\di
S_{\eta,t}}$ at the boundary of $\calB_t$, or, more specifically along the
edges of the $S_{\eta,t}$, the pieces of the surfaces $S_\lam$ lying in
$\Sg_t$. The second term in (\ref{DgEdot_expression2}) vanishes because
$e_i$ is covariantly constant on the 2-surfaces $S_\lam$, and thus on
$S_{\eta,t}$.

Equivalently, the second term in (\ref{DgEdot_expression1}) can be shown to
vanish in a more elementary, if less picturesque, way using the constancy of
$e_i$ and (\ref{a0_w}),
which implies that $\ag_b \sg_{0c}\eg^{abc} = 0$ because $\tilw^a = 0$.
Similarly the characteristics of the surviving contribution to
$\Dg\dot{\tilE}$,
$e_i\:\eg^{abc}\di_b\bar{\sg}_{0c}$, can be derived from (\ref{w_evolution}).
Since $\dot{\tilw}$ vanishes inside $\calB_t$ and $\bar{\sg}_{0c}$ vanishes
outside $\tilde{u}^a \equiv 2\eg^{abc}\di_b\bar{\sg}_{0c}$ lives on the
boundary
of $\calB_t$. Moreover $\di_a \tilde{u}^a = 0$, so $\Dg\dot{\tilE}_i^a =
e_i\ \tilde{u}^a$ shows, like our previous analysis, that $\Dg H_b$ generates
the birth of $\tilE$ field lines along the boundary of $\calB_t$.

$\Dg H_b$ also generates an entirely analogous evolution of the $\tilB$ field
that ensures that along with the $\tilE$ field lines are born corresponding
$\tilB$ field lines so that (\ref{generalized_SD}) is satisfied.
The $H_l$ term in $H_b$ then generates a 3-diffeomorphism that moves the
field lines (initially) away from $\di \calB_t$.

In summary, the $\tilE$ and $\tilB$ fields evolve as follows in the 2-sphere
solution: Until birth begins $\tilE$ and $\tilB$ are zero. During birth
loops of $\tilE$ and $\tilB$ flux, i.e. field lines, are generated by $\Dg
H_b$.
As they are created these field lines move out, forming a torus in space once
the birth is completed. This torus expands and recontracts, and then the events
of the birth are repeated in reverse during death, leaving ultimately $\tilE =
\tilB = 0$.

In a generic slicing a given $S_\lam$ and $\Sg_t$ will be tangent only if
$t_{min}$ (or $t_{max}$) of $S_\lam$ coincides with $t$, and then only at one
point.
In other words, the `disks'
$S_{\eta,t}$ are points in a generic slicing.
The union, over $\lam_1$ and $\lam_2$ of these points of tangency to a fixed
$\Sg_t$ then forms a line $\ell$ in $\Sg_t$.

I have used Ashtekar's Hamiltonian density (which is correct when $\Sg_{i\,0a}$
is of the lapse-shift form) outside $\calB_t$ to emphasize that the evolution
there is the same as in Ashtekar's theory. This approach becomes confusing
when the $S_{\eta,t}$ are points. It is then better to treat all of $\Sg$
uniformly
by using the Hamiltonian density $\tilde{\cal H}_\Sg = -A^i_0 D_a \tilE^a_i -
\Sg_{i\,0a}\,[\tilB^{i\,a} - \phi^{ij}\tilE^a_j]$ everywhere, treating
$\Sg_{i\,0a}$ as independent of $A$ and $\tilE$ in the Poisson brackets,
and only afterward substituting the particular form of $\Sg_{i\,0a}$ into the
resulting evolution equation. The occurrence of births and deaths is then
indicated by $\Sg_{i\,0a} \neq 0$ at some points where $\tilE = 0$.\footnote{
In the generic case, in which births occur only along the line $\ell$,
$\Sg_{i\,0a} = \frac{1}{2}\eg_{abc}\tilE^b_i N^c$ everywhere except on $\ell$.
Moreover, when the $S_\lam$ are smooth in the coordinates adapted to the
slicing
and threading $\Sg_{i\,0a}$ is smooth, so $\Sgauge$ on $\ell$ is the limit
of $\Sgauge$ off $\ell$.

This suggests that births could be incorporated in the Ashtekar theory if only
certain singular $\vec{N}$ were allowed. In fact this cannot be done in a
straightforward way, since the evolution of $\tilE$ generated by $H_l$,
$\dot{\tilE}^a_i = \pounds_{\vec{N}}\ \tilE^a_i$, preserves $\tilE=0$ for
any $\vec{N}$.}

We can conclude that the crucial feature of the new Hamiltonian of Section
\ref{three_plus_one} which lets it, unlike Ashtekar's Hamiltonian, generate
births and deaths is the presence of $\tilE$ independent contributions
\be
\Dg\tilde{\cal H}_b = \beta_{i\,a}\tilB^{i\,a},
\ee
where $\bg_{i\,a}$ may be non-zero when $\tilE = 0$. The Ashtekar Hamiltonian
contains only terms that are positive powers of $\tilE$, so
$\{\tilE^a_i, H_{Ash}\} = 0$ when $\tilE= 0$. $H_{Ash}$ can only generate
changes in $\tilE$ in the support of $\tilE$ or its boundary.


\section*{Acknowledgments}

The work described here was started while I was at the Department of Physics of
Washington University,
in St. Louis, supported by NSF grant PHY 92-22902 and NASA grant NAGW 3874.
I thank Clifford Will for his patience. The work was completed at
Utrecht University. There I thank my hosts Bernard de Wit and Gerard 't Hooft
and the other members of the Institute for Theoretical Physics for their
hospitality. The present paper was submitted during a visit to the University
of Vienna, where I thank my hosts Peter Aichelburg and Robert Beig.

Several people stimulated my thinking with questions and insights.
Many aspects of the present work have benefited substantially from discussions
with Ingemar Bengtsson. Furthermore, Letterio Gatto, J. J. Duistermaat, and
H. Urbantke helped me via discussions of diffeomorphisms, boundary value
problems,
and the geometry of self-dual 2-forms, respectively.

Finally, Xiao Feng Cai, Carlo Rovelli, Lee Smolin,
and Jan Smit provided essential encouragement.


\appendix
\section{Faraday lines as the classical limit of graph basis states}
\label{classical_loop}

In this Appendix it is shown that, in the classical limit, the graph basis
state $|\cg,j\rangle$ associated with a graph consisting of an intersection
free loop $\cg$, carrying spin $j$, essentially represents an isolated line
of $\tilE$ flux along $\cg$. (The result trivially generalizes to the graph
basis states corresponding to disjoint collections of such loops.)

More precisely, $|\cg,j\rangle$ can be written as the sum of two states
$|\cg,j\rangle^+$ and $|\cg,j\rangle^-$ such that, in the connection
representation (see \cite{Ashtekar_Tate}) in which the operator $\hat{A}^i_a$
acts by multiplication,
\bearr
\langle A|\hat{\tilE}^a_i |\cg,j\rangle^+ & = & \Egj^a_i\loopstate^+ + O(\hbar)
		\label{eigenvalue1}	\\
\langle A|\hat{\tilE}^a_i |\cg,j\rangle^- & = & -\Egj^a_i\loopstate^- +
O(\hbar),
		\label{eigenvalue2}
\eearr
with
\be
\Egj^a_i = e_i\: \Dgthree^a_\cg,		\label{E_eigenvalue}
\ee
$\Dgthree^a_\cg(x) = \int_\cg \dg^3(x - z) dz^a$, and $e_i = j\hbar n_i$,
where the unit $SO(3)$ vector $n_i$ is covariantly constant on $\cg$.
Unless, that is, the holonomy of $A$ around $\cg$ is $\One$.
Notice that when the holonomy is not $\One$ all covariantly constant
vectors on $\cg$ are constant multiples of $n$, so $\Egj$ is uniquely
defined, up to sign, by $\cg$, $j$, and $A$.

As $\hbar$ becomes small $\langle A|\hat{\tilE}^a_i |\cg,j\rangle^\pm
\rightarrow 0$ unless $j \sim O(1/\hbar)$, in which case $e_i$ is
finite and (\ref{E_eigenvalue}) is an isolated Faraday line carrying
flux $e_i$.
Thus, according to our claim $|\cg,j\rangle$ in fact represents
two Faraday lines, of opposite flux, along $\cg$.

Let's prove the claim (\ref{eigenvalue1}), (\ref{eigenvalue2}).

In the connection representation the graph basis state $|\cg,j\rangle$
is represented by \cite{Reisenberger94}
\be
\loopstate = tr H^{(j)}[A,\cg],
\ee
(times a normalization $\sqrt{2j+1}$ which will be dropped here).
$H^{(j)}[A,\cg] = \calP\:exp(\int_\cg A^i J_i^{(j)})$ is the spin $j$
holonomy around $\cg$, and the $J_i^{(j)}$ are the spin $j$ representations
of the antihermitian $su(2)$ generators. In other words $\loopstate$ is
the spin $j$ Wilson loop.

Now the result (\ref{eigenvalue1}), (\ref{eigenvalue2}) can be derived by
straightforward mathematics.
The holonomy referred to the base point $p \in \cg$ can be written as
\be
H^{(j)}_p [A,\cg] = e^{\theta^i(p) J_i^{(j)}}.
\ee
Its trace (which is independent of $p$) is
\bearr
tr [e^{\theta^i(p) J_i^{(j)}}] & = & \sum_{m = -j}^j e^{i|\theta|m}\\
	& = & \frac{\sin(j + \frac{1}{2})|\theta|}{\sin \frac{1}{2}|\theta|} \\
	& = & c(|\theta|) e^{i|\theta|j} + c(|\theta|)^* e^{-i|\theta|j},
\eearr
where $c(|\theta|) = \frac{1}{2} [ 1 - i \cot \frac{|\theta|}{2} ]$.

It is easy to show that $\theta^i(p)$ is covariantly constant on $\cg$,
and that its magnitude is $|\theta| = \int_\cg n_i A^i$, where $n_i$ is
the $SO(3)$ unit vector $\theta_i/|\theta|$.
Thus
\be
\loopstate = c(|\theta|)e^{i\int_\cg n_i A^i j} + c(|\theta|)^* e^{-i\int_\cg
n_i A^i j}.
\ee

Taking
\bearr
\loopstate^+ & = & c(|\theta|)e^{i\int_\cg n_i A^i j} \\
\loopstate^- & = & c(|\theta|)^* e^{-i\int_\cg n_i A^i j}
\eearr
we find (\ref{eigenvalue1}) and (\ref{eigenvalue2}):
\bearr
\langle A |\hat{\tilE}^a_i|\cg,j\rangle^+ & \equiv &
-i\hbar\frac{\dg}{\dg A^i_a}\loopstate^+			\\
	& = & [ j + \frac{\di}{\di |\theta|} \log c(|\theta|)] \hbar n_i
		\int_\cg \dg^3(x - z) dz^a \loopstate^+		\\
	& = & \Egj^a_i \loopstate^+  + O(\hbar).
\eearr
(\ref{eigenvalue2}) follows similarly.

$\Egj^a_i = e_i \Dgthree^a_\cg$ might not seem like a proper eigenvalue
field, even in the classical limit, because $e_i$ depends on $A$, the
argument of $\loopstate$. However, by a gauge transformation one can
always make $n_i = \dg^3_i$ on all of $\cg$, leading to
\be
\Egj^a_i = {}_0\Egj^a_i \equiv j\hbar\dg^3_i \Dgthree^a_\cg,
\ee
which is manifestly independent of $A$. While the components of
$\Egj$ depend on gauge, they do not depend on the gauge equivalence class
of the $A$ field, which is the true argument of the gauge invariant functions
$\loopstate^\pm$. Hence, in a suitable gauge fixing $|\cg,j\rangle^\pm$ is,
when $\hbar\rightarrow 0$, an eigenstate of
$\hat{\tilE}^a_i$ with eigenvalue $ \pm {}_0\Egj^a_i$. So $|\cg,j\rangle^\pm$
represent, in the classical limit, Faraday lines, which are described in an
arbitrary gauge by $\tilE^a_i = \pm \Egj$.

\section{Tetrads from bases of self-dual 2-forms}\label{sig_tet}

A proof is given of a somewhat elaborated form of a theorem of Capovilla, Dell,
Jacobson
and Mason \cite{CDJM}.
The very efficient proof given in \cite{CDJM} relies on spinorial techniques.
Here $SO(3)$ tensors are used.

\noindent {\bf Theorem}
\bearr
\Sg_i\wedge\Sg_j = \frac{1}{3}\dg_{ij}\Sg_k\wedge\Sg^k 	\label{hyp1} \\
\Sg_i\wedge\Sg^i \neq 0,       		\label{hyp2}
\eearr
1), implies that there exists a non-singular co-tetrad $e^I{}_\m$ such that
\be  \label{claim}
\Sg_i = \frac{1}{2} e^0\wedge e^i + \frac{1}{4}\eg_{ijk} e^j\wedge e^k.
\ee
Furthermore,

\noindent 2), the metric $g_{\m\n} = e^I{}_\m\dg_{IJ}e^J{}_\n$ is uniquely
determined
by $\Sg_i$, and is in fact equal to the Urbantke metric \cite{Urbantke83}
defined by
\be
\sqrt{g} g_{\m\n} = 4
\Sg_{1\,\m\ag}\Sg_{2\,\bg\cg}\Sg_{3\,\dg\n}\eg^{\ag\bg\cg\dg}.
\ee

\noindent 3), $e_0{}^\m$ may be chosen to be any unit vector of $g$ ($e_I{}^\m$
is the inverse
of $e^I{}_\m$), but once this vector is chosen $e^I{}_\m$ is uniquely
determined. Equivalently,
$e^I{}_\m$ is unique up to the action of the $SO(3)_R$ subgroup of $SO(4)$ on
the internal
index $I$.

\noindent 4), When $\Sg_i$ is real there exist either real $e^I{}_\m$
satisfying (\ref{claim}),
corresponding to a positive definite metric, or pure imaginary $e^I{}_\m$,
corresponding
to a negative definite metric.

\noindent {\bf Proof}:

First let's establish 1) by constructing a cotetrad $e^I{}_\m$ satisfying
(\ref{claim}).
Any vector $t^\m$ allows us to define three 1-forms
\be		\label{fi}
f^i{}_\m = 2 t^\sg\Sg_{i\,\sg\m}.
\ee
These will ultimately, after a rescaling, serve as the `spatial' part,
$e^i{}_\m$ of $e^I{}_\m$.

When $t$ and the $\Sg_i$ are real the $f^i$ are easily shown to be linearly
independent:
if they were not there would exist non-zero, real $a_i$ such that $a_i f^i =
0$. This implies
$t^\sg[a^i\Sg_i]_{\sg\m} = 0$, which means that $rank[a^i\Sg_i] = 2$ or $0$,
implying, in turn,
that
\be			\label{a2Sig2}
0 = a^i\Sg_i\wedge a^j\Sg_j = a^i a^j\frac{1}{3}\dg_{ij}\Sg_k\wedge\Sg^k.
\ee
Since $\Sg_k\wedge\Sg^k \neq 0$ this requires $\dg_{ij}a^i a^j = 0 \Rightarrow
a^i = 0$.
That is, the $f^i$ are linearly independent.

If the $\Sg_i$ and/or $t$ are complex the situation is less simple. The $f^i$
may then be
complex and linear dependence requires only the existence of non-zero {\em
complex} $a^i$
such that $a_i f^i = 0$. Now $\dg_{ij}a^i a^j = 0$ has non-zero complex
solutions.
By (\ref{a2Sig2}), if $a^i$ is such a solution $rank[a^i\Sg_i] = 2$.
($rank[a^i\Sg_i] = 0$,
i.e. $a^i\Sg_i = 0$, is excluded because this would imply $0 =
a^i\Sg_i\wedge\Sg_j
= a_j(\frac{1}{3}\Sg_k\wedge\Sg^k) \rightarrow a_j = 0$.)
If $t$ is chosen to be a null vector of $a^i\Sg_i$ then the corresponding $f^i$
will not
be linearly independent.

Nevertheless, it will now be shown that there {\em are} $t$ with $f^i$ that are
independent.
Let $N_3 \subset \Complex^3$  be the set of solutions to $\dg_{ij}a^i a^j = 0$,
and let
$N_4 = \{t \in \Complex^4 | \exists a \in N_3 \suchthat t^\m [a^i\Sg_i]_{\m\n}
= 0 \}$.
$N_4$ is the set of $t$ having linearly dependent $f^i$. $N_4$ can be thought
of as the
subset of $\Complex^4$ swept out by the two dimensional null plane of
$a^i\Sg_i$ as $a$
ranges over $N_3$. $N_3$ is a two dimensional cone in $\Complex^3$, but one of
these dimensions
corresponds to rescalings of $a$, which do not affect the null plane of
$a^i\Sg_i$.
Thus $N_4$ has, at most, dimension $2 + 1 = 3$. This makes it a lower
dimensional subset of
$\Complex^4$ so $\Complex^4 - N_4$ is not empty. It will be shown below that
$N_4$ is the
null cone of the metric induced by $\Sg_i$.

Choose any $t \in \Complex^4 - N_4$. The corresponding $f^i$ are linearly
independent.
Furthermore, $t^\m f^i{}_\m = 0$ so, if $\ag$ is a 1-form such that $t^\m\ag_\m
= 1$,
$\{\ag, f^i\}$ form a basis of 1-forms, which can be used to expand the
$\Sg_i$. Taking
(\ref{fi}) into account
\be	\label{Sg_expansion1}
\Sg_i = \frac{1}{2}\ag\wedge f^i + \bg_{ijk} f^j\wedge f^k,
\ee
with $\bg_{ijk}$ antisymmetric in $j\:k$.

Using this expansion we may write
\bearr
\Sg_i\wedge\Sg_j & = & \frac{1}{2}\bg_{jlm} \ag\wedge f^i\wedge f^l\wedge f^m +
					\frac{1}{2}\bg_{ilm} f^l\wedge f^m\wedge \ag\wedge f^j \\
				 & = & \eg^{lm}{}_{(i}\bg_{j)lm} \ag\wedge f^1\wedge f^2\wedge f^3.
\eearr
(\ref{hyp1}) then implies $\eg^{lm}{}_{(i}\bg_{j)lm} \propto \dg_{ij}$, that
is,
\be
\eg^{lm}{}_i \bg_{jlm} = \kappa \dg_{ij} + \eta^k\eg_{ijk}
\ee
for some $\kappa \neq 0$ and $\eta^i$ ((\ref{hyp2}) implies $\kappa \neq 0$).
Now, since $\bg_{ijk}$ is antisymmetric in $j\:k$,
\be
\bg_{ijk} = \frac{1}{2}\eg^n{}_{jk}[\eg^{lm}{}_n\bg_{ilm}] =
\frac{1}{2}\kappa\eg_{ijk}
+ \dg_{i[j}\eta_{k]}.
\ee
Hence,
\bearr
\Sg_i & = & \frac{1}{2}[\ag - 2\eta_k f^k]\wedge f^i + \frac{1}{2}\kappa
\eg_{ijk} f^j\wedge f^k \\
	& = & 2\kappa [ \frac{1}{2} f^0\wedge f^i + \frac{1}{4}\eg_{ijk} f^j f^k],
\label{Sg_expansion2}
\eearr
with $f^0 = \frac{1}{2\kappa}(\ag - 2\eta_k f^k)$. To arrive at the form
(\ref{claim}) of
$\Sg_i$ it is simply necessary to absorb the factor $2\kappa$ in a rescaling of
the basis 1-forms:
let
\be
e^I{}_\m = \sqrt{2\kappa} f^I{}_\m
\ee
then
\be
\Sg_i = \frac{1}{2} e^0\wedge e^i + \frac{1}{4}\eg_{ijk} e^j\wedge e^k.
\ee
This proves 1).

Now for 2). $e^I{}_\m$ determines a spacetime metric $g_{\m\n} = e^I{}_\m
\dg_{IJ} s^J{}_\n$.
The formulas
\bearr
[det\ e] g_{\m\n} & = & 4
\Sg_{1\,\m\ag}\Sg_{2\,\bg\cg}\Sg_{3\,\dg\n}\eg^{\ag\bg\cg\dg} \\
det\ e & = & \frac{1}{6}\Sg_{i\,\ag\bg}\Sg^i_{\cg\dg}\eg^{\ag\bg\cg\dg} \neq 0
\eearr
can be verified by substituting the expansion (\ref{claim}) for the $\Sg_i$.
Together they
show that $g_{\m\n}$ is determined by the $\Sg_i$.

3). How unique is $e^I{}_\m$? First note that $e^I{}_\m$ is the $f^I{}_\m$
corresponding
to a normalized $t^\m$, namely $t_0^\m = \sqrt{2\kappa} t^\m$. Since $t_0^\m
e^I{}_\m = \dg^I_0$
$t_0^\m = e_0{}^\m$, the `time' element of $e_I{}^\m$, the inverse $e^I{}_\m$.

Now let's suppose we are given a cotetrad $e'^I{}_\m$ satisfying (\ref{claim}).
Are there
distinct $e^I{}_\m$, also satisfying (\ref{claim}), with $e_0{}^\m$ pointing in
the same
direction as $e'_0{}^\m$? Take $t^\m = \lam e'_0{}^\m$ with $\lam > 0$.
\be	\label{fi_given_e}
f^i{}_\m = 2 t^\sg \Sg_{i\,\sg\m} = \lam e'^i{}_\m.
\ee
These $f^i$ are linearly independent, so $t \in \Complex^4 - N_4$. Thus $\Sg_i$
can be expanded
according
to (\ref{Sg_expansion2}) in $f^I\wedge f^J$ and according to (\ref{claim}) in
$e'I\wedge e'^J$.
Making use of (\ref{fi_given_e}) we find
\be
2\kappa[\frac{1}{2} f^0\wedge \lam e'^i + \frac{1}{4}\lam^2\eg_{ijk} e'^j\wedge
e'^k]
 =  \frac{1}{2}e'^0\wedge e'^j + \frac{1}{4}\eg_{ijk}e'^j\wedge e'^k
\ee
The linear independence of the $e'I\wedge e'^J$ now demands $2\kappa =
1/\lam^2$ and
$\frac{1}{\lam} f^0\wedge e'^i = e'^0\wedge e'^i$ - which implies $f^0 = \lam
e'^0$.
Hence
\be
e^I{}_\m = \sqrt{2\kappa} f^I{}_\m = \frac{1}{\lam}\lam e'I{}_\m = e'I{}_\m.
\ee
The cotetrads $e^I{}_\m$ satisfying (\ref{claim}) are thus in one to one
correspondence
with the rays of vectors $t \in \Complex^4 - N_4$.

Note that $e_0{}^\m$ cannot lie in the null cone ${\cal C} \subset \Complex^4$
of the
metric $g$, since $e^0{}_\m = g_{\m\n}e_0{}^\n$ has to be a unit vector and
finite.
Thus ${\cal C} \subset N_4$.

Note also that $\Sg_i = [e\wedge e]^{+i}$, a self-dual component of $e^I\wedge
e^J$. The
$\Sg_i$ are therefore invariant under anti-self-dual $SO(4)$ (i.e. $SO(3)_R$)
transformations on the internal index $I$ of $e^I{}_\m$. It follows from the
anti-self-duality
of the generators of $SO(3)_R$ that $SO(3)_R$ transformations consist of an
$SO(4)$
boost by an arbitrary rapidity $\theta^i$, accompanied by a spatial rotation by
an angle
$|\theta|$ about $\theta^i$. Hence $e_0{}^\m$ can be boosted to be parallel to
any given
unit vector $t_0^\m$, provided the rest of the tetrad is rotated apropriately.
(Note
that $e^I \rightarrow - e^I$ is in $SO(3)_R$, so $SO(3)_R$ will take $e_0$ from
the
past to the future unit norm shell in the Lorentzian section, though via a path
that departs
from this section at intermediate points). We see that there is an $e^I{}_\m$
corresponding to each
$t \in \Complex^4 - {\cal C}$, so $N_4 = {\cal C}$. Furthermore, we see that
all $e^I{}_\m$
are $SO(3)_R$ transforms of one $e^I{}_\m$. Thus the freedom in $e^I{}_\m$ for
given
$\Sg_i$ is precisely $SO(3)_R$.

4), the case of real $\Sg_i$.

If $t^\m$ is taken to be real then the $f^i$ are real, $\kappa$ is real, and,
if $\ag$
is taken to be real, $\eta^i$ is real. Hence the $f^I$ are real in this case.
However, $\kappa$
can be positive or negative, leading to either real $e^I$ and a positive
definite metric,
or pure imaginary $e^I$ and a negative definite metric.\QED

\section{The case $rank \tilE = 2$} \label{rank2app}

In this appendix it is shown that when $\tilE$ is degenerate but of rank $2$
Ashtekar's constraints
are still equivalent to the new constraints (\ref{Gauss_law}) and
(\ref{generalized_SD}), and $\Sgauge$ is of the lapse-shift form
(\ref{lapse_shift_form}).

When $rank\tilE = 2$ both the internal span of $\tilE^a_i$ (the span of
$\tilE^a_i$ seen as three
internal vectors labeled by $a$), and its external span, are two dimensional.
Therefore there
exists an internal vector $v^i$ and an external vector $w_a$ such that
$v^i\tilE_i^a = w_a\tilE_i^a
= 0$. Note also that $\tilde{\varepsilon}^i_a = \frac{1}{2} \epsilon^{ijk}
\epsilon_{abc}
\tilE^b_j \tilE^c_k \propto v^i w_a$, and is non-zero.

Let's first show that Ashtekar's constraints (\ref{Gauss_law}), (\ref{vector}),
and (\ref{scalar})
imply the full constraints, (\ref{Gauss_law}) and (\ref{generalized_SD}). The
Gauss law constraint
(\ref{Gauss_law}) is the same in both sets, so it remains only to show that the
vector
and scalar constraints, (\ref{vector}) and (\ref{scalar}), imply
(\ref{generalized_SD}).

The vector constraint implies
\be \label{vec2}
\eg_{ijk} \tilB^{i\,a} \tilep^j_a = 0,
\ee
and the scalar constraint can be written as
\be \label{scal2}
\dg_{ij}\tilB^{i\,a} \tilep^j_a = 0.
\ee
(\ref{vec2}) and (\ref{scal2}) require that $\psi^{ij} = \tilB^{i\,a}
\tilep^j_a$ be trace free
and symmetric. On the other hand $\psi^{ij} \propto \tilB^{i\,a} w_a v^j$, so
it is rank $1$.
Together these requirements imply that $\psi^{ij} = 0$, which in turn implies
that
$\tilB^{i\,a}w_a = 0$. In other words, $\tilB^{i\,a}$ is in the external span
of $\tilE$:
\be
\tilB^{i\,a} = \theta^{ij} \tilE^a_j  \label{linear_dep}
\ee
for some $\theta$. Substituting (\ref{linear_dep}) back into the vector
constraint gives
\be
0 = \theta^{ij}\tilE^a_j\tilE^b_i \eg_{abc},
\ee
i.e. the component of $\theta$ acting in the internal span of $\tilE$ is
symmetric.
{}From (\ref{linear_dep}) it can be seen that the remaining components of
$\theta$ may be
chosen freely. $\theta$ may, therefore, be chosen trace free and symmetric.
The constraint (\ref{generalized_SD}),
\be
\exists \phi^{ij} \mbox{trace free, symmetric} \suchthat
\tilde{B}^{i\,a} - \phi^{ij} \tilde{E}^a_j = 0,
\ee
is thus satisfied.

Now let's show that the general solution to the field equation
(\ref{Sigma_constraint}),
\be
\Sg_{(i\,0a} \tilE^a_{j)} \propto \dg_{ij},    \label{feq1.2}
\ee
is the lapse, shift form of $\Sg_{i\,0a}$:
\be  \label{lapse_shift_form2}
\Sg_{i\,0a} =
\frac{1}{4} N \eg_i{}^{jk}\eg_{abc} \tilE^b_j \tilE^c_k +
\frac{1}{2}\eg_{abc}\tilE^b_i N^c.
\ee
(\ref{feq1.2}) always implies that
\be \label{feq1.2_imp}
\Sg_{i\,0a}\tilE^a_j = \tilde{c}_1 \dg_{ij} + \eg_{ikj} \tilde{c}_2^k,
\ee
for some $\tilde{c}_1$ and $\tilde{c}_2$. When $rank\tilE = 2$
(\ref{feq1.2_imp}) can be
simplified as follows. Contracting (\ref{feq1.2_imp}) with $v^j$ gives $0 =
\tilde{c}_1 v_i
+ \eg_{ikj}\tilde{c}_2^k v^j$, which implies $\tilde{c}_1 = 0$ and $\tilde{c}_2
\parallel v$.
Thus
\be  \label{rank2form}
\Sg_{i\,0a}\tilE^a_j \propto \eg_{ikj} v^k.
\ee
The component of $\Sg_{i\,0a}$ in the external span of $\tilE$ has only one
degree of freedom.
The component transverse to the external span of $\tilE$ is, of course, totally
unconstrained
by (\ref{feq1.2}), and so has three degrees of freedom. This makes for a total
of four degrees
of freedom, which is exactly what the lapse, shift form
(\ref{lapse_shift_form2}) of $\Sg_{i\,0a}$
has, a good sign.

(\ref{rank2form}) may be written as
\be
\Sg_{i\,0a}\tilE^a_j = \frac{1}{4}\eg_{ikj}
[\eg^{klm}\eg_{abc}\tilE^b_l\tilE^c_m] N^a
= \frac{1}{2} \eg_{abc}\tilE^b_i N^c \tilE^a_j,
\ee
in which only the component of $N$ along $w$ contributes. $\Sg_{i\,0a}$ itself
is thus of the form
\be \label{rank2form_2}
\Sg_{i\,0a} = \frac{1}{2} \eg_{abc}\tilE^b_i N^c + c_{3\,i} w_a.
\ee
$c_{3\,i}$ captures the three degrees of freedom of the component of
$\Sg_{i\,0a}$ transverse
to the external span of $\tilE$. Note, however, that the components of $N$
transverse to $w$
contribute to (\ref{rank2form_2}), and that their contribution spans
expressions of the form
$c_i w_a$ with $c_i v^i = 0$. Thus $c_3$ may be set parallel to $v$, making the
last term in
(\ref{rank2form_2}) proportional to $v_i w_a \propto \eg_i{}^{jk}\eg_{abc}
\tilE^b_j \tilE^c_k$.
(\ref{rank2form_2}) is then of the form (\ref{lapse_shift_form2}).

\section{Characteristic distributions} \label{characteristic_dist}

Characteristic distributions can be defined for any $n$ dimensional
submanifold, $M$, of a $d$
dimensional space:
\be
\Dg_M^{\m_1...\m_n} = \int_M \dg^d(x - z) dz^{[\m_1}...dz^{\m_n]}.
\ee
These distributions have the important property that for an arbitrary $n$-form,
$g$,
\be
\int_M g = \int_{\Real^d} \Dg_M^{\m_1...\m_n} g_{\m_1...\m_n} d^d x.
\ee

The $\Dg$'s of submanifolds and those of their boundaries are connected by the
identity
\be  \label{boundary_id}
\di_{\m_1} \Dg_M^{\m_1\m_2...\m_n} = - \frac{1}{n} \Dg_{\di M}^{\m_2...\m_n}.
\ee
(\ref{boundary_id}) follows directly from Stokes theorem: for an arbitrary
$n-1$-form $f$
\bearr
\int_{\Real^d} \di_{\m_1} \Dg_M^{\m_1\m_2...\m_n} f_{\m_2...\m_n} d^d x & = &
			- \int_{\Real^d} \Dg_M^{\m_1...\m_n} \di_{[\m_1}f_{\m_2...\m_n]} d^d x \\
		& = & - \frac{1}{n} \int_M d \wedge f \\
		& = & - \frac{1}{n} \int_{\di M} f \\
		& = & - \frac{1}{n}\int_{\Real^d} \Dg_{\di M}^{\m_2...\m_n}f_{\m_2...\m_n}
d^d x,
\eearr
the arbitrariness of $f$ implying (\ref{boundary_id}).


\end{document}